\documentclass[fleqn,usenatbib]{mnras}
\usepackage[T1]{fontenc}
\usepackage{ae,aecompl}
\usepackage{threeparttable}

\usepackage{graphicx}	% Including figure files
\usepackage{amsmath}	% Advanced maths commands
\usepackage{amssymb}	% Extra maths symbols

\newcommand{\rem}{{\rm rem}}
\newcommand{\recy}{{\rm recy}}

\newcommand{\bh}{{\rm BH}}
\newcommand{\Ms}{~{\rm M}_\odot}

\newcommand{\nbodysix}{\texttt{NBODY6}~}
\newcommand{\SSEET}{\texttt{SSE18}~}
\newcommand{\BSEET}{\texttt{BSE18}~}
\newcommand{\SSEVN}{\texttt{SSEVN}~}

\title[BHs spins and masses to infer GW sources history]{Using final black hole spins and masses to infer the formation history of the observed population of gravitational wave sources}

\author[M. Arca Sedda and M. Benacquista]{
M. Arca Sedda,$^{1}$\thanks{E-mail: m.arcasedda@ari.uni-heidelberg.de}
M. Benacquista,$^{2,3}$
\\
$^{1}$Zentrum f\"{u}r Astronomie der Universit\"{a}t Heidelberg, Astronomisches Rechen-Institut, M\"onchhofstr. 12-14, 69120 Heidelberg\\
$^{2}$Center for Gravitational Wave Astronomy, University  of Texas Rio Grande Valley, One University Blvd,  Brownsville TX 78520,  USA\\
$^{3}$Division of Astronomy, National Science Foundation, 2415 Eisenhower Ave, Alexandria, VA 22314, USA
}

\date{Accepted XXX. Received YYY; in original form ZZZ}

\pubyear{2018}

\begin{document}
\label{firstpage}
\pagerange{\pageref{firstpage}--\pageref{lastpage}}
\maketitle

\begin{abstract}
In this paper we propose a novel technique to constrain the progenitor binary black hole (BBH) formation history using the remnant masses and spins of merged black holes (BHs). Exploring different models, we found that dynamically formed BBHs are distributed differently in the mass-spin plane than those that have formed in isolation. Stellar evolution recipes crucially affect the remnant mass distribution, suggesting that future efforts should be devoted to finding a common way of modelling the evolutionary phases of single and binary stars. Our simple approach has allowed us to place weak constraints on the origin of the presently observed population of merged BHs, although with high uncertainties. 
Our results show that the fingerprints of different BBH formation channels will emerge as soon as LIGO detects more than $\sim 10^2$ merger events. This work provides a way of thinking that can be easily used both by people working on isolated and on dynamical BBH formation and evolution.
\end{abstract}


%%%%%%%%%%%%%%%%%%%%%%%%%%%%%%%%%%%%%%%%%%%%%%%%%%
\section{Introduction}

The recent detections of gravitational waves (GWs) achieved by the LIGO/VIRGO consortium have, for the first time, allowed the observation of the last phases preceding the merger of stellar-mass binary black holes (BBH) \citep{abbott16a, abbott16b, abbott17a, abbott17b}. 

These are the first direct detections of BBHs to date, opening the window on a new era for astronomy as a whole, further pushed by the recent first observation of a double neutron star merger \citep{abbott17d}.

One of the puzzles coming from the LIGO/VIRGO detections is the origin of stellar BHs with masses above $20\Ms$. 
Indeed, before the detection of GW150914 \citep{abbott16a}, most BHs known resided in binary systems while accreting from their companion. The X-ray emission from the accretion disk around the BH allowed for the detection of $\sim 22$ BHs so far \citep{xbinrev,casares17}. 
These BHs have masses in the range $1.6-18\Ms$ \citep{casares17}. In contrast, the LIGO sources have masses of $10-36\Ms$, with merger products as heavy as $62 \Ms$.

These GW sources may have originated either in the field---from isolated binary stellar evolution \citep{belczynski08,belczynski16b,belczynski17} or in a triple system \citep{antonini17}---or through dynamical interactions in dense young massive clusters \citep{zwart00,mapelli16,banerjee16,banerjee17}, globular clusters, \citep{downing10,rodriguez15,rodriguez16,rodriguez17,askar17} or galactic nuclei \citep{bartos17,antonini16b,OLeary16,hoang18,ASCD17b,ASG18}.
Different environments would affect differently the BBH properties. For instance, if a BBH merges while orbiting a supermassive black hole, it may appear heavier to GW detectors, since its rapid motion around the supermassive black hole can induce a noticeable Doppler shift, which in turn can increase the observed BBH mass \cite{Chen17,ASCD17b}.
Driving a BBH toward coalescence by dynamical scattering is not very easy. The most likely channel for this to occur is through repeated binary-single or binary-binary interactions, which can lead the BBH to progressively shrink and eventually be ejected from the cluster and merge later, or merge promptly inside the inner cluster regions \citep{rodriguez17,samsing17}. These interactions can also lead to the formation of a hierarchical system, in which the inner BBH merger is driven by Kozai-Lidov oscillations \citep{antonini12,naoz13,naoz16} or to a short-lived, non-hierarchical triple in which the outer component induces a rapid increase in the eccentricity of the inner binary, accelerating the merging process \citep{ASKLI18}.
While all the scenarios above explain the observationally inferred BBH merger rate quite well, it is still unclear whether the high masses observed by LIGO/VIRGO can tell us something about single \citep{spera15} or binary stellar evolution physics \citep{belczynski16b}, or if they are due to repeated collisions among BHs retained in their parent clusters after a ``first generation'' of merging events \citep{rodriguez17}. In the latter scenario, post-merger BHs are retained in the parent stellar cluster and undergo subsequent mergers, increasing their mass up to the presently observed values. 

The current sensitivity of the three detectors allows us to obtain crucial information on the BHs masses before and after the merger, to infer the distance at which the merger took place, and to get an estimate of the pre- and post-merger spin of the BHs involved.
Whether the initial spin and masses can be connected to the BHs formation history is currently matter of strong debate, both on the isolated binary evolution \citep{belczynski16b,belczynski17}, and the dynamical \citep{morawski18} sides.
Indeed, understanding whether different formation scenarios and physical processes might leave a fingerprint in merging BBHs observed properties is one of the key questions of modern GW astronomy. For instance, BHs forming through multiple mergers are expected to have larger masses and higher spins compared to BHs born from the first merger generation \citep{gerosa17,fishbach17a}.
Providing a robust strategy to unveil how spotting the signs of different channels has also important implications for improving our understanding of stellar evolution for single and binary stars. For instance, the observed distribution of BH chirp masses can be used to infer information on the BH natal kick distribution \citep{zevin17}, thus allowing a big step forward in the comprehension of how BHs form and evolve. With the number of detections increasing as LIGO and Virgo reach full sensitivity it will be also possible to determine what is the most likely state for BHs spin, either aligned or isotropically distributed \citep{stevenson17,talbot17,farr17}.

Disentangling the formation pathways of BBHs from GW observations is a many-faceted problem. It requires a self-consistent and simultaneous treatment of star cluster evolution (to account for dynamically formed BBHs), single and binary stellar evolution (to properly calculate the BHs nascent mass and spin), general relativity (needed to calculate the final BH mass and spins), and cosmology (crucial to infer the metallicity distribution and the total number of detectable mergers in the local Universe).

In this paper, we propose a general framework that allows us to explore how different formation channels can affect the distribution of the final mass and spins of merged BHs, taking into account the most recent results in the field of dense cluster dynamics, isolated binary evolution and numerical relativity at the same time.

Using this framework, we show that our approach can be used to assess which is the most probable scenario for the formation of the observed GW sources, discussing also how this can be used to predict the future population of observed BHs once the number of detections becomes statistically significant.

The paper is organized as follows: in section 2, we describe our models; in section 3, we describe the detection likelihood for merger remnants  for each model;  in section 4, we discuss likely formation scenarios for the known  gravitational-wave mergers; section 5 is devoted to summarize our main conclusions. 

\section{Creating mock BBHs for field and dynamical scenarios}

To determine whether different BBHs formation channels can lead to differences observable in the remnant mass and spin distribution, we created two samples of BBHs assuming that they formed either in the field or through dynamical interactions in a dense stellar environment.

Many other parameters must be taken into account to provide a complete overview of the possible outcomes of BBH evolution. We can divide these crucial parameters into two classes: \textit{stellar} and \textit{environmental}. 
Stellar parameters depend only on the physics of stellar evolution adopted, which basically come down to the selection of stellar evolution recipes and the choice of the stellar metallicity. Environmental parameters, instead, depend upon the physical processes that drive the BBH formation, i.e. the formation channel (isolated or dynamical). These include the BHs spin alignment, the BBH mass ratio and the probability for a merged BBH to undergo a new merger. 

In this section, we describe how the BH natal masses and spins (Sect. \ref{natal}) are calculated in our models, how different BHs are paired in BBH (Sect. \ref{pair}), and how we obtained the remnant mass and spin (Sect. \ref{remnant}).

\subsection{BH natal masses and spins}
\label{natal}

In all our models, we selected stars with initial masses in the range $18-150\Ms$ according to a \cite{kroupa01} mass function. We produce 7000 binaries for each model investigated, using the same number for either isolated or dynamical channels. We chose this number since it corresponds to the number of objects in a Kroupa distribution with masses above $18\Ms$ that are expected to form in a stellar system with a total initial mass of $\sim 2.6\times 10^6\Ms$.

For dynamically formed binaries, we can expect the component BHs to have formed independently. Therefore, we can determine the final BH masses for these stars using the latest version of the package \texttt{SSE} \citep{hurley00} which includes the stellar wind prescriptions of \cite{belczynski10} and FeNi core mass calculations adapted from \cite{belczynski08}. We extracted this version from the direct N-body code \nbodysix\footnote{\url{https://www.ast.cam.ac.uk/~sverre/web/pages/nbody.htm}} \citep{aarseth12}, where it is currently implemented. With these updates, a star with an initial mass of $100\Ms$ will lead to a BH of mass $M_{\rm BH} = 40\Ms$, assuming a metallicity of $Z=2\times 10^-4$. For the remainder of this paper, we will refer to this code as \SSEET.

For isolated binary stellar evolution, we note that a general consesus is not yet established. There are several codes that follow the evolution of a binary star, all characterized by different underlying assumption and physical processes. Among these, we note here \texttt{StarTrack} \citep{belczynski08}, \texttt{MOBSE} \citep{giacobbo18}, \texttt{MESA} \citep{paxton11}, and \texttt{COMPAS} \citep{barrett18}. All these codes can successfully reproduce a population of merging BBHs similar to the observed population. However, with the same initial conditions they can lead to different BBH semi-major axis, mass, and eccentricity, all depending on the assumptions being made. A detailed comparison among all these codes is beyond the scope of this paper, we will instead use two different approaches to characterize our isolated BBH population.

For the first approach we use the most recently updated version of the \texttt{BSE} code for binary stellar evolution \citep{hurley02}, also extracted from \nbodysix. Throughout the paper, we will refer to this code as \BSEET. We will use \BSEET as a standalone code that models the binary evolution from its initial phase down to the final BBH configuration. We retain only the systems that will merge in a Hubble time. We refer to this as the B model.

In the second approach, we explore the effects of different stellar evolution models using a simplified model for the BBH population. In addition to the \SSEET models, we also use an alternative model for isolated stellar evolution based on \texttt{PARSEC}, the so-called Padova stellar tracks \citep{bressan12}. We will explain the details of this approach in the next section. This approach will be refered to as the I model.

\begin{figure}
\centering
\includegraphics[width=8cm]{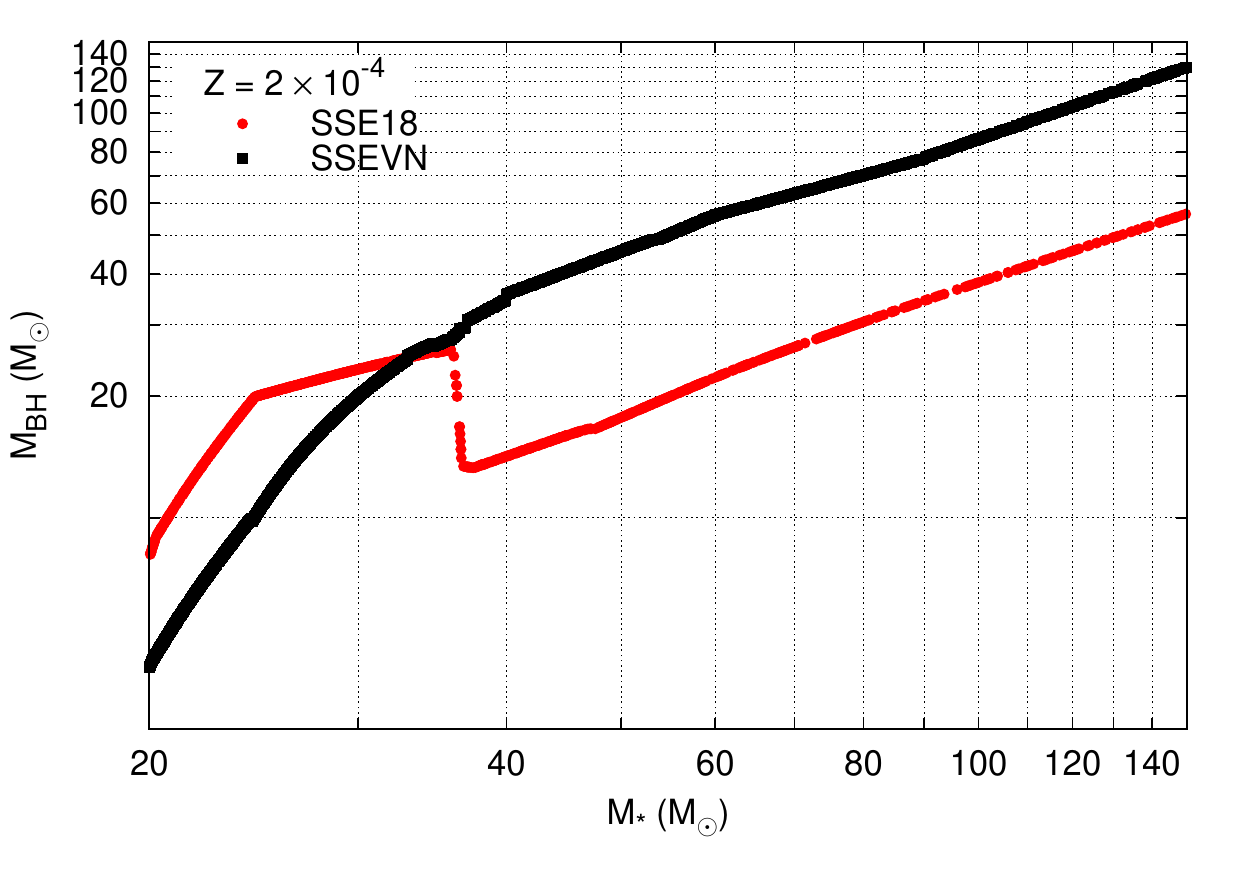}
\includegraphics[width=8cm]{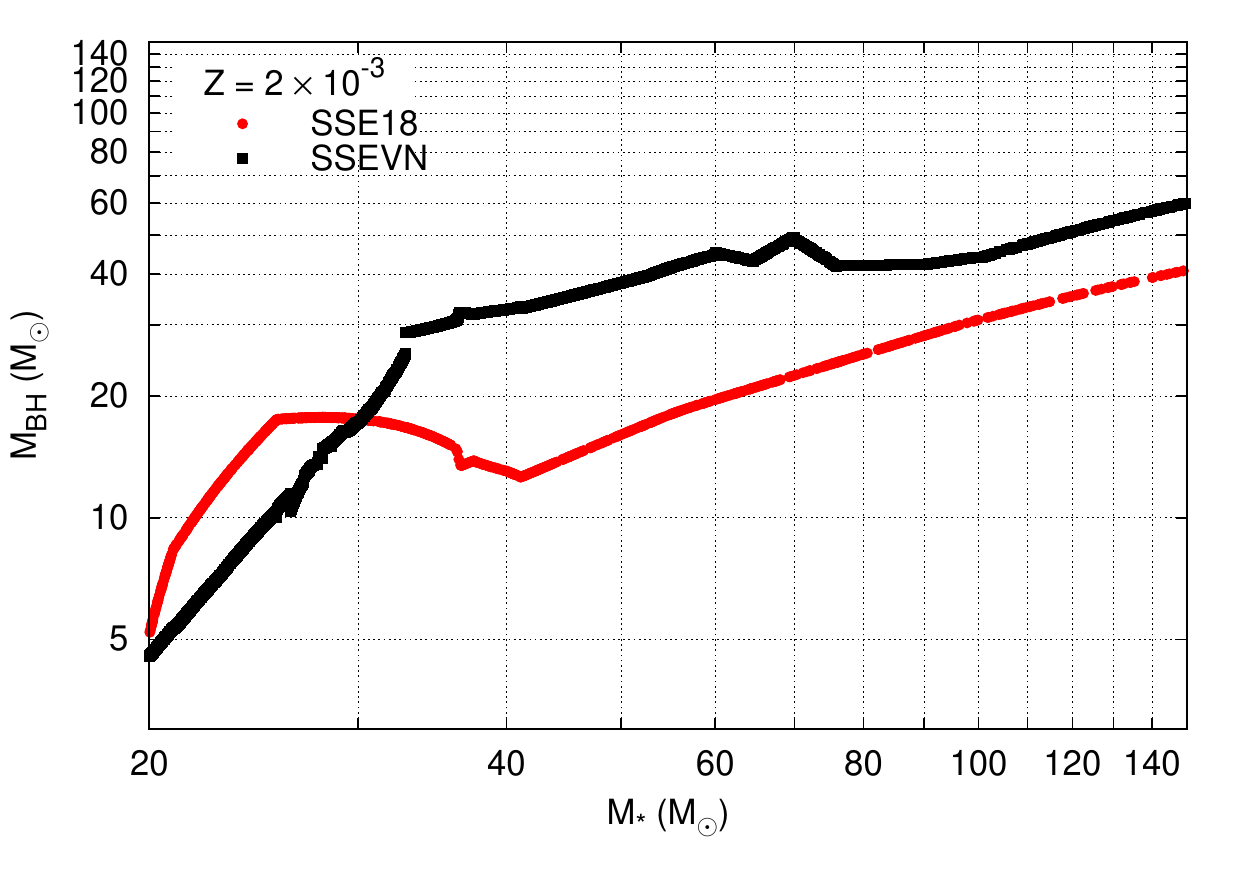}
\includegraphics[width=8cm]{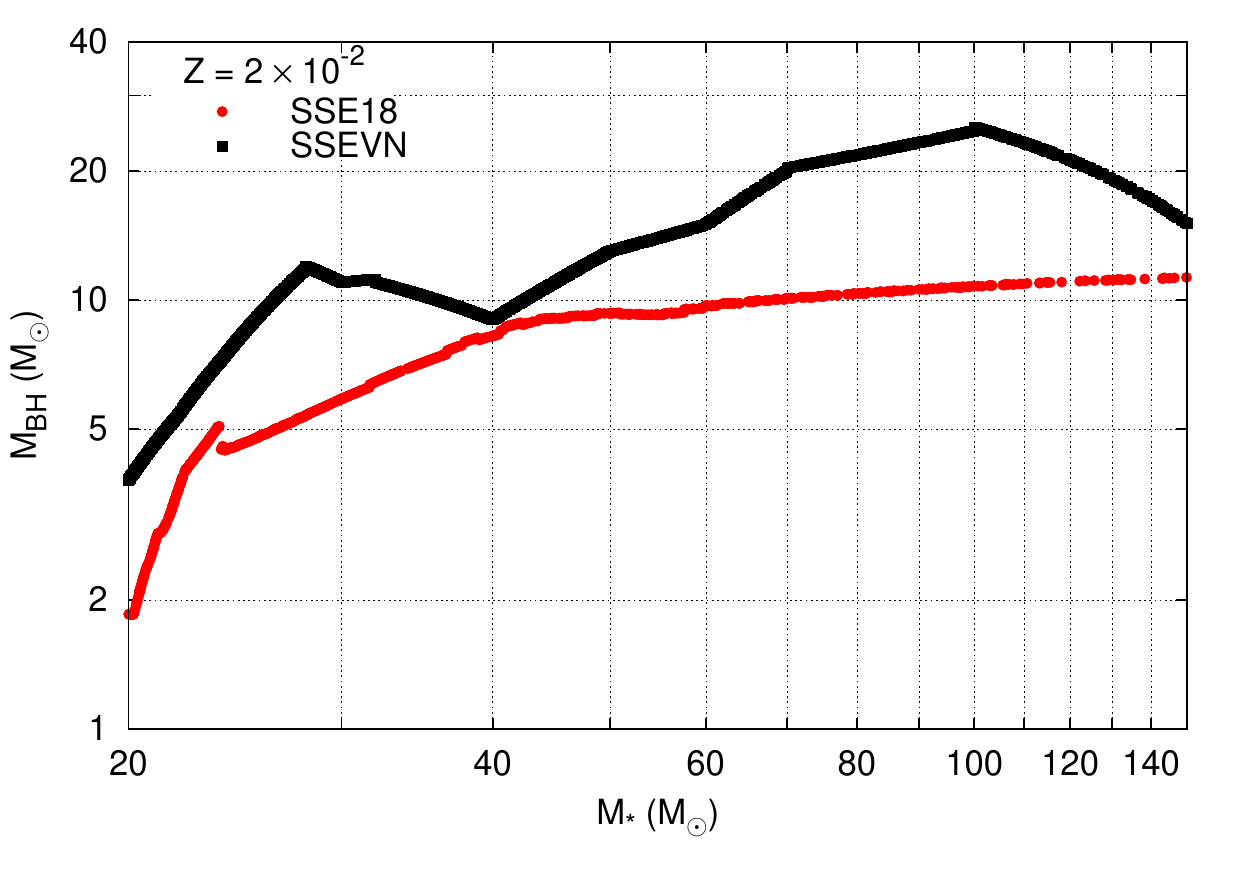}
\caption{Final BH mass as a function of the progenitor initial mass as calculated by the updated \texttt{SSE} code (filled red circle) and the \citet{spera15} formulae for different metallicities: $Z = 2\times 10^{-4}$ (top panel),$2\times 10^{-3}$ (central panel), $2\times 10^{-2}$ (bottom panel).
(filled black squares).}
\label{mbh}
\end{figure}

Using PARSEC, \citet{spera15} showed that the BH population originating from metal-poor stars can have masses that are substantially larger than previously thought. Consequently, we adapted their results to compare their BH mass spectrum with the standard output obtained from the \SSEET code. In Figure~\ref{mbh}, we show the mass spectra for both models, assuming metallicities of $Z=2\times 10{-4},~2\times 10^{-3}$, and $2\times 10^{-2}$. Note, for example, that a metal poor star with initial mass $M_*=100\Ms$, will turn into a BH with final mass $M_{\rm BH} \simeq 85\Ms$---more than 2 times larger than the value obtained through the \SSEET recipes.

Using a handful of fitting formulae to recover the results of \citet{spera15} (see Appendix for the case with $Z = 0.0002$), we can quickly explore how a different stellar evolution recipe can affect the formation and evolution of BBHs and the also the properties of the post-merger BH. For the sake of clarity, we will refer to these fitting formulae as \SSEVN\footnote{The original stellar evolution code developed by \cite{spera15}, called \texttt{SEVN}, can be downloaded at \url{http://web.pd.astro.it/mapelli/group.html\#software}}.

To determine the BH spins, we follow \citet{belczynski17} and assume that the natal spin depends on the progenitor carbon oxygen core mass, $M_{\rm CO}$, through the relation
\begin{equation}
a_\bh = 
\begin{cases}
0.85            & M_{\rm CO} < m_1,      \\
k_1M_{\rm CO}+k_2 & m_1 < M_{\rm CO} < m_2, \\
a_{\rm low}     & M_{\rm CO} \geq m_2, \\
\end{cases}
\label{EQanat}
\end{equation}
The parameters $k_1$, $k_2$, $a_{low}$, as well as $m_1$ and $m_2$ depend on the stellar metallicity, as discussed in \citet{belczynski17}.

We note that the BH natal spin can be directly connected with the BH mass through the fitting relation
\begin{eqnarray}
a_\bh(M_\bh) = \frac{A}{\left[1+\left(M_\bh/M_0\right)^B\right]^C}.
\label{inspin}
\end{eqnarray}
which is found from combining all the \citet{belczynski17}. To reduce the error introduced in evaluating the fitting parameters, we fixed $M_0 = 60\Ms$ and $B=8$, and then obtained $A=0.89\pm 0.02$ and $C=0.21 \pm 0.02$ from the fit.

\begin{figure}
\centering
\includegraphics[width=8cm]{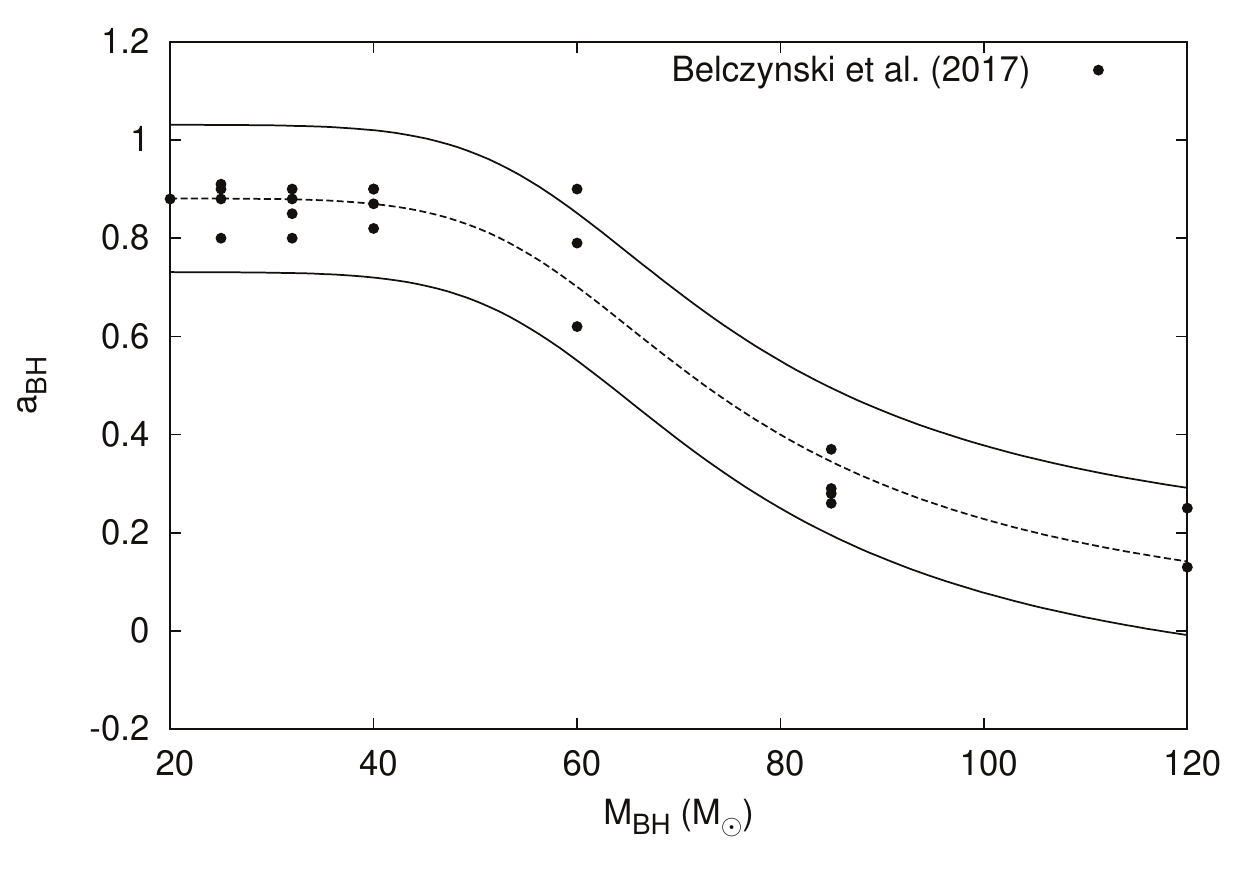}
\caption{BH natal spin as a function of the BH mass. Filled black circles are data from \citet{belczynski17}.}
\label{bspin}
\end{figure}

For model I, we obtain the mass for BHs in the following way: first we select the progenitor star mass from the \citet{kroupa01} mass function, and then using either the \SSEET code or the \SSEVN fitting formulae we determine $M_{\rm BH}$ for those that are sufficiently massive to turn into black holes. For the spins, we associate each BH with an initial spin that is randomly chosen between $a_{\rm BH}(M_{\rm BH}) \pm 0.15$ as shown in Figure~\ref{bspin}. These two limiting curves contain the \citet{belczynski17} dataset.

For model B, the BH masses and spins are generated directly as output of the \BSEET code and calculated during the different stages of the evolution of the binary. However, if a component progenitor evolves without being affected by its companion, \BSEET assigns a constant value for the natal spin. For the sake of coherence, we replace the constant value with a randomly chosen spin as in model I when this occurs.

\subsection{Coupling BHs in binaries}
\label{pair}

Once the masses and spins have been chosen, we then need to determine the binary properties based on their environmental parameters.

In order to study the ``field'' BBHs, we created two model sets based on the I and B models. As described above, we select the masses from the \citet{kroupa01} model and assign spins appropriately. The alignment of the spins is chosen so that the spins are either aligned or mildly misaligned, in order to explore the role of spin alignment on the final BH properties.

The I models are the least realistic of our models. They have been chosen to represent the limiting case of isolated formation in which binary stellar evolution processes are negligible. In this case each BBH progenitor is considered a detached system where common envelope and tidal circularization effects can be ignored. Therefore, we can assume that every binary in this model will merge within a Hubble time and the masses of the component BHs are simply chosen through either \SSEET or \SSEVN.

We incorporate the effects of binary evolution in our B models. The LIGO discoveries have motivated a number of new tools for modeling binary stellar evolution \citep{marchant16,demink16,belczynski16b,belczynski17,giacobbo18}. However, most of these are not freely available for public use. For this reason, we adapted \BSEET for use in the B models. We leave a comparison between \BSEET and other stellar population synthesis tools to a future work. 
As discussed in several recent works, different assumptions on the various mechanisms involved in binary stellar evolution can affect significantly the merging BBHs properties. For instance, the assumption of a dependence between BHs natal kicks and the fall-back mass can maximize the differences between isolated and dynamical BBHs \citep{zevin17}. Supernovae explosion mechanisms play a significant role in determining the merging BBHs final properties \citep{giacobbo18} and the survival of double neutron stars \citep{kruckow18}, while the common envelope phase, which is still poorly understood \citep{ivanova13}, may significantly affect the BBH merger rate \citep{kruckow18}. Rather than providing a parameter space study, we restrict our analysis to a handful of physically motivated parameters, as discussed in the next section.  

Although a detailed description of all the features included in \BSEET is out of the scope of this paper, we summarize here the main updates related to stellar BH physics and the default choices that we have made for our models.
\begin{itemize}
\item {\bf Natal kick prescription.} The user can decide how the linear velocity imparted to the nascent BH is calculated. Kicks are extracted according to either 
a flat distribution or Maxwellian distribution with dispersion $\sigma$. The user can decide both the flat distribution limiting values and the Maxwellian velocity dispersion. Two prescriptions are available for kick amplitudes: the ``full'' prescription, which assumes that BHs kick are equal to the full kick imparted to neutron stars, or the ``fall-back'' prescription, according to which the BH nascent kick is a fraction of neutron star kicks and the fraction depends on the stellar core mass, $V_{\rm kick}^{\rm BH} = (1-f_{\rm flbck})V_{\rm kick}^{\rm NS}$. Further options that can be enabled are related to neutron stars and white dwarfs natal kicks.  
\item {\bf Mass loss prescription.} The code allows to select different treatments for mass loss: i) the \cite{hurley00} basic rates, ii) the SSE standard treatment plus a treatment for luminous blue variables, iii) the \cite{belczynski10} updates for stellar winds of massive stars \citep{vink01,vink05}, iv)  the \cite{belczynski10} updates without the bi-stability jump. 
\item {\bf Common envelope.} A few minor updates have been included in the treatment of common envelope phase: i) the binary binding energy parameter $\lambda$ is forced to be 0.5 for naked stars, ii) in case of eccentric binaries the orbital energy is reduced by a factor that depends on the pericentral distance. 
\end{itemize}

We have chosen for our defaults to have
\begin{itemize}
\item {\bf Natal kick prescription.} We chose a Maxwellian distribution with $\sigma = 100~{\rm km/s}$ with a fallback scaling: $V_{\rm BH} \propto \left(1-f_{\rm flbck}\right)$. These low kick have been chosen to explain several observations of x-ray binaries \citep{willems05, zuo14, wong14, belczynski16b, wisocki18}.
\item {\bf Mass loss prescription.} We adopted the mass loss and stellar wind according the \citet{belczynski10} update.
\item {\bf Common envelope.} We use a common envelope efficiency factor $\alpha_{\rm CE}=0.1$ and a binding energy parameter $\lambda_{\rm CE} = 1$, motivated by massive stars \citep{xu10,loveridge11,giacobbo18}.
\end{itemize}

\begin{figure}
\includegraphics[width=8cm]{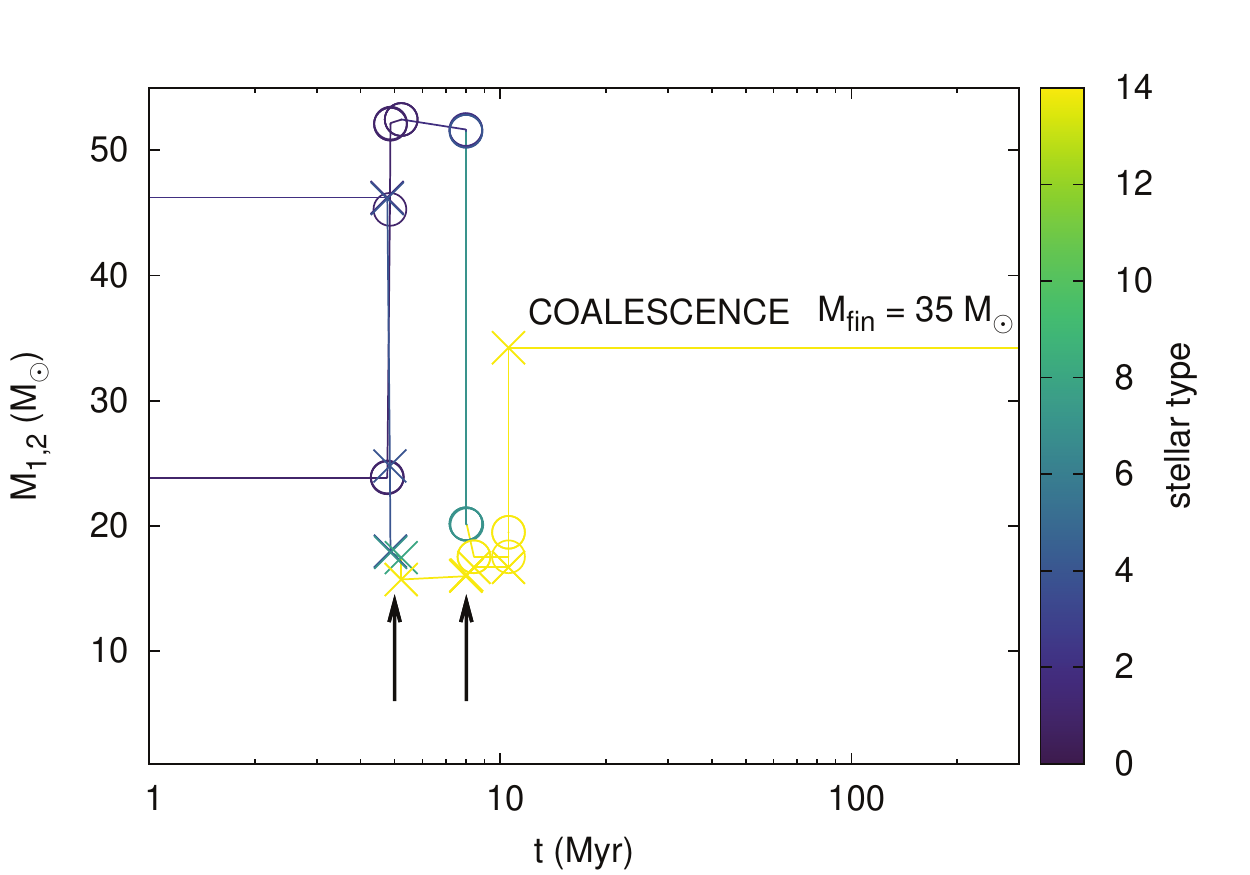}\\
\includegraphics[width=8cm]{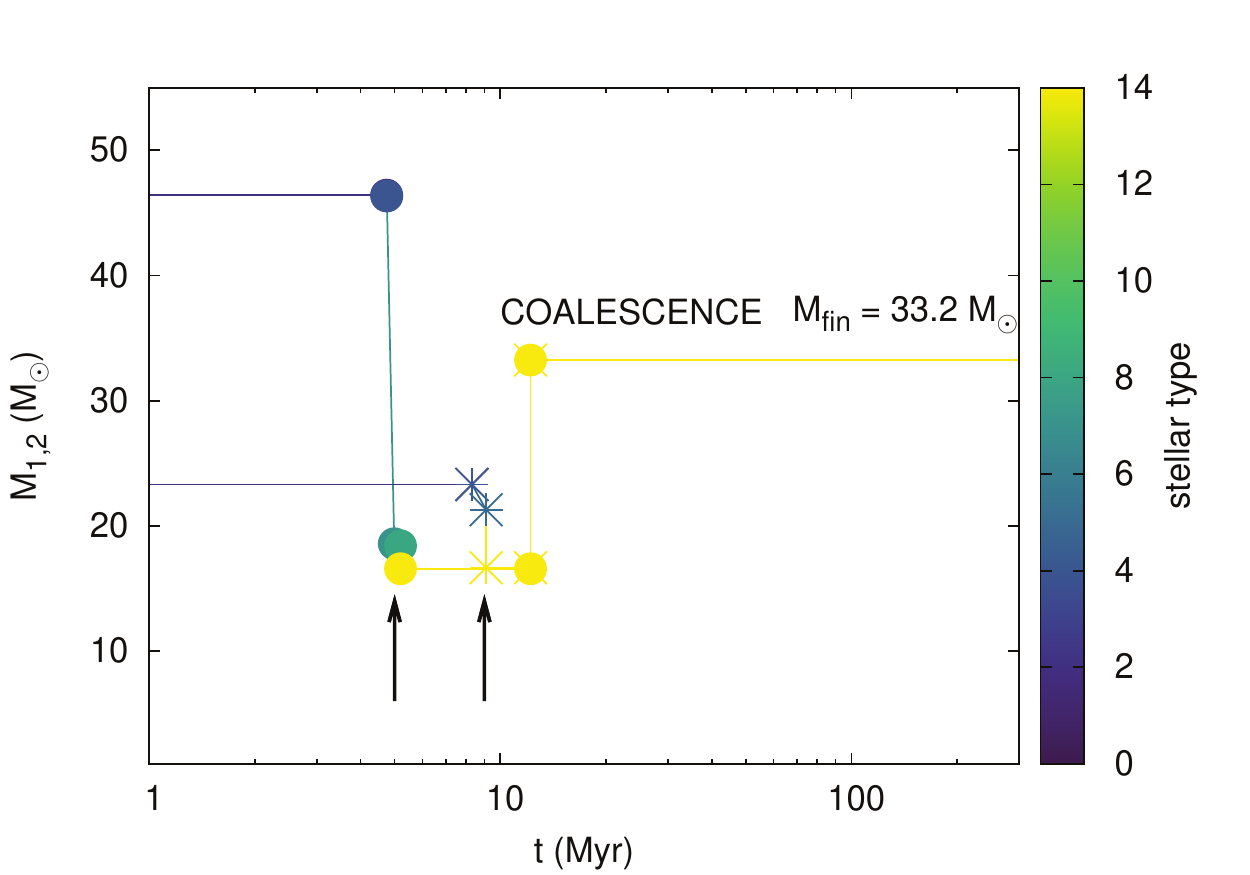}
\caption{Mass evolution for a binary star with initial masses $M_1 = 23.4\Ms$, $M_2 = 46.9\Ms$, semi-major axis $a_{\rm semi} = 133$ AU and initial eccentricity $e = 0.42$ evolved with the \BSEET package (top panel) and calculated for the I model, i. e. using \SSEET (bottom panel). The color coding, taken from \citet{hurley00}, marks different evolutionary stages. At $\sim 12$ Myr, the two BHs merge.}
\label{ex}
\end{figure}

As an example of the comparison between the I model and the B model, we follow the typical example of the binary evolution in \BSEET and its equivalent in \SSEET. Consider two stars with initial masses $M_1=23.4\Ms$ and $M_2 = 46.9\Ms$, with initial eccentricity $e_0 = 0.41$ and orbital period $T=17601$ days. We evolve this system through \BSEET and then consider the case where both components evolve in isolation with \SSEET. The evolution of the system in both cases is shown in Figure~\ref{ex}.

In the B models, the primary fills its Roche lobe in the ``Hertzsprung gap'' phase and starts feeding the secondary, reaching the core-He burning stage. It then loses its envelope, appearing as a naked He main sequence, further evolving as a naked Hertzsprung gab star and turning into a BH. Later, the former secondary fills its Roche lobe and begins accreting onto the BH, evolving toward the core-He burning stage. At this point, the common envelope phase takes place. The binary shrinkage is very efficient during this phase and the separation of the components is reduced by a factor of $\sim 30$. The secondary then evolves from the naked He main sequence star phase to BH without any further interaction with its companion. The resulting BBH has a total mass $M_1+M_2 = 35\Ms$.

In the much simpler I model evolution, we assume that the two stars pass through all the stellar stages until turning into BHs, leading to the formation of a BBH with a total mass $M_1+M_2 = 33.2\Ms$. In this particular case, we see that the interaction of the binary components does not play a significant role in determining the BBH final mass since the difference between the models is within $10\%$.

Although \BSEET will give the spins of the component BHs, it does not give their alignments. Thus, we explore the spin alignment in both I and B models in an ad hoc fashion. We assume that spins of isolated BBH may have a degree of alignment. Defining the angle $\theta$ to be the projection of the spin onto the orbital angular momentum, we select values of $\theta$ from the distribution
\begin{equation}
P(\cos\theta){\rm d}\cos\theta = \left(\cos\theta+1\right)^{n_\theta}{\rm d}\cos\theta ,
\label{ntheta}
\end{equation}
for the spin of each component BH in the binary. The parameter $n_\theta$ is used to set the degree of alignment, so that a larger value of $n_\theta$ corresponds to more aligned distribution of the spins. In our models, we choose $n_\theta = \infty,~2,~{\rm or}~0$, corresponding to fully aligned, thermal, or flat distributions, repsectively.

The complete set of parameter choices for both I and B models for the formation of isolated BBH is listed in Table~\ref{tab1}.

Binaries formed through dynamical interactions require a different set of parameters. The easiest way to create mock ``dynamical'' BBHs is to simply pair BHs with randomly chosen masses and spins. However, although their formation is determined through chaotic interactions, several simulations have pointed out that dynamically formed merging BBHs are characterised by relatively high mass ratios i.e. comparable masses (see for instance \cite{amaro16}).

In their extensive work based on Monte-Carlo simulations, \citet{rodriguez16} showed that dynamically formed BBH with sufficiently low merger time-scales are usually characterized by mass ratios above $0.4$, with a tiny probability for merging systems with $0.2<q<0.4$ (see their Figure 9). They also showed that the BBH mass ratio does not on average depend on the total mass.

In order to reflect this aspect, we paired BH masses in our reference model so that they have a mass ratio $q = M_1/M_2 \geq 0.5$. The masses are generated through either \SSEET or \SSEVN. In the event that the BBH mass ratio is not robustly constrained, we also carry out a few models assuming a flat distribution for this parameter. To discriminate between these models, we identify the minimum mass ratio $q_{\rm min}$, such that $q > q_{\rm min}$ for all BBHs within a given model. The orientation and magnitude of spins, on the other hand, are distributed randomly according to Equation~\ref{inspin}.

As a further parameter to be considered, we also take into account the possibility that some merged BBHs can be retained within their parent cluster. These systems may then pair with another BH in the cluster and undergo a second merger. \citet{rodriguez17} recently suggested that around half of the BBH merger events in globular clusters can occur before the BBH ejection, and in a substantial fraction of cases the post-merged BH is retained. In order to investigate how a {\em second generation} of merger events can change the distribution of $M_{\rm rem}-a_{\rm rem}$, we define an additional parameter $f_\recy = 0.1$ that describes the fraction of merged BHs that can re-interact with other BHs in the cluster.

Table \ref{tab1} summarizes the main properties of our models. We refer to ``reference model'', I0, for isolated binaries  as the one with fully aligned spins ($n_{\theta} = \infty$), low-metallicity ($Z = 2\times 10^{-4}$) and \SSEET stellar evolution. A similar setup characterizes the reference model for isolated binaries evolved with \BSEET, namely B0.
The corresponding ``reference model'' for dynamical BBH, D0, is characterised by a minimum mass ratio $q_{\rm min}=0.5$ and the same low-metallicity. Moreover, we assumed that BBHs undergo only one merger event ($f_\recy = 0$).

\begin{figure}
\centering
\includegraphics[width=8cm]{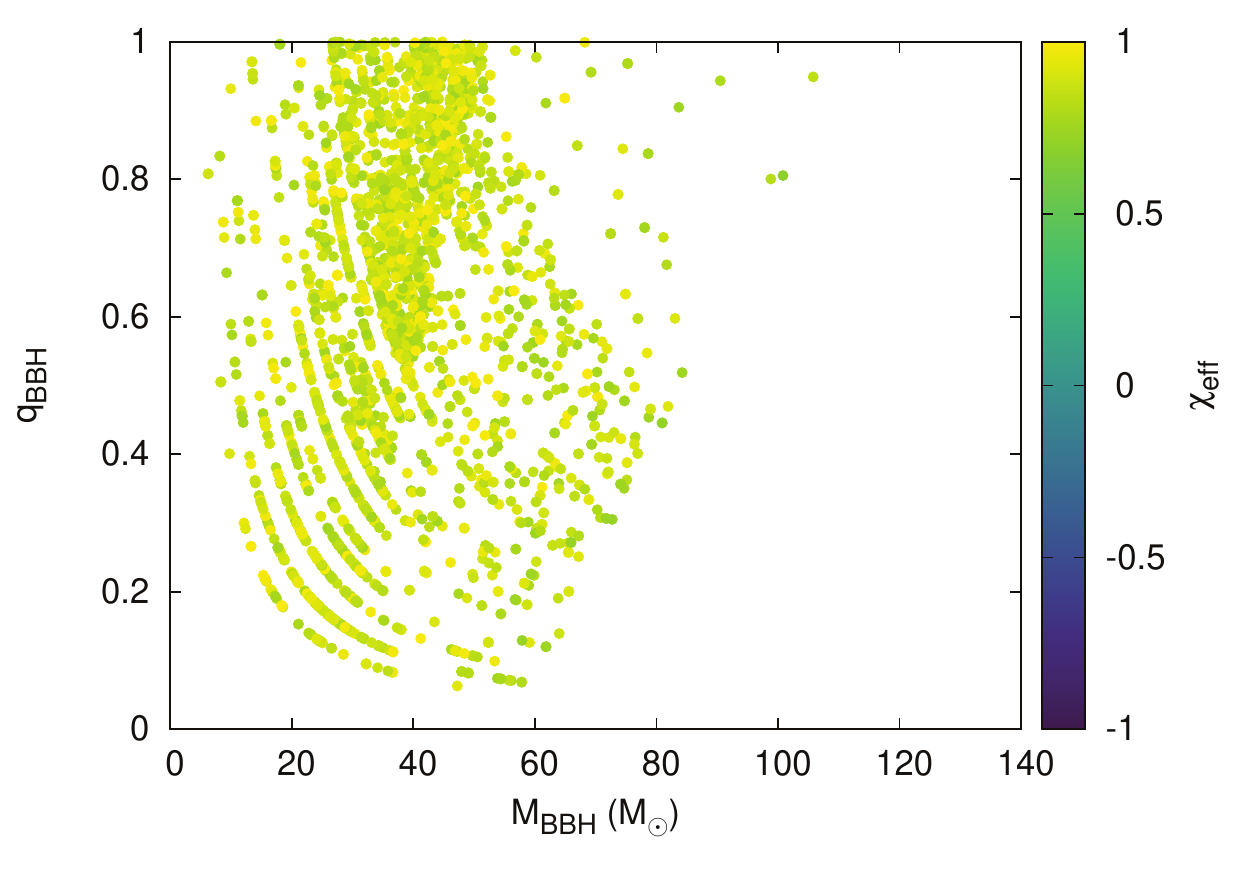}\\
\includegraphics[width=8cm]{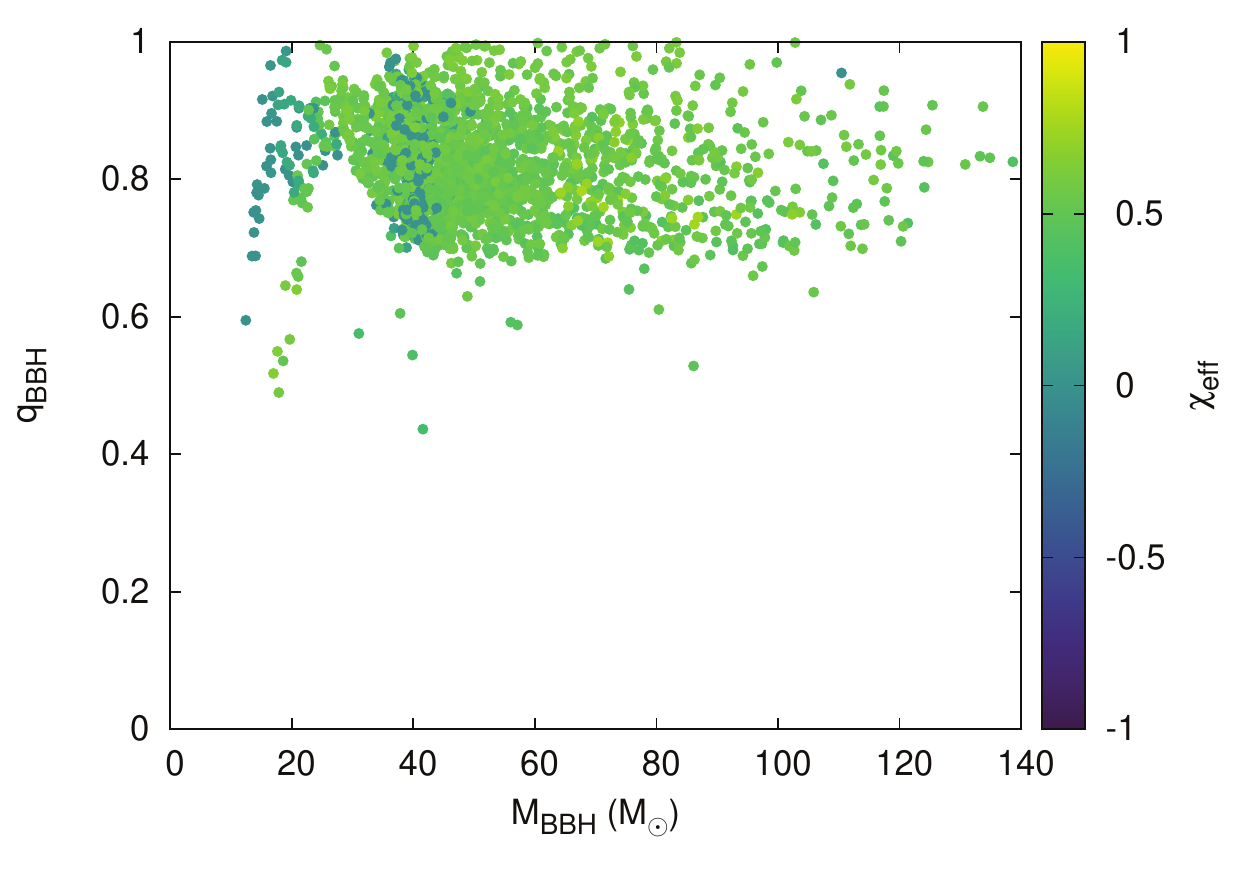}\\
\includegraphics[width=8cm]{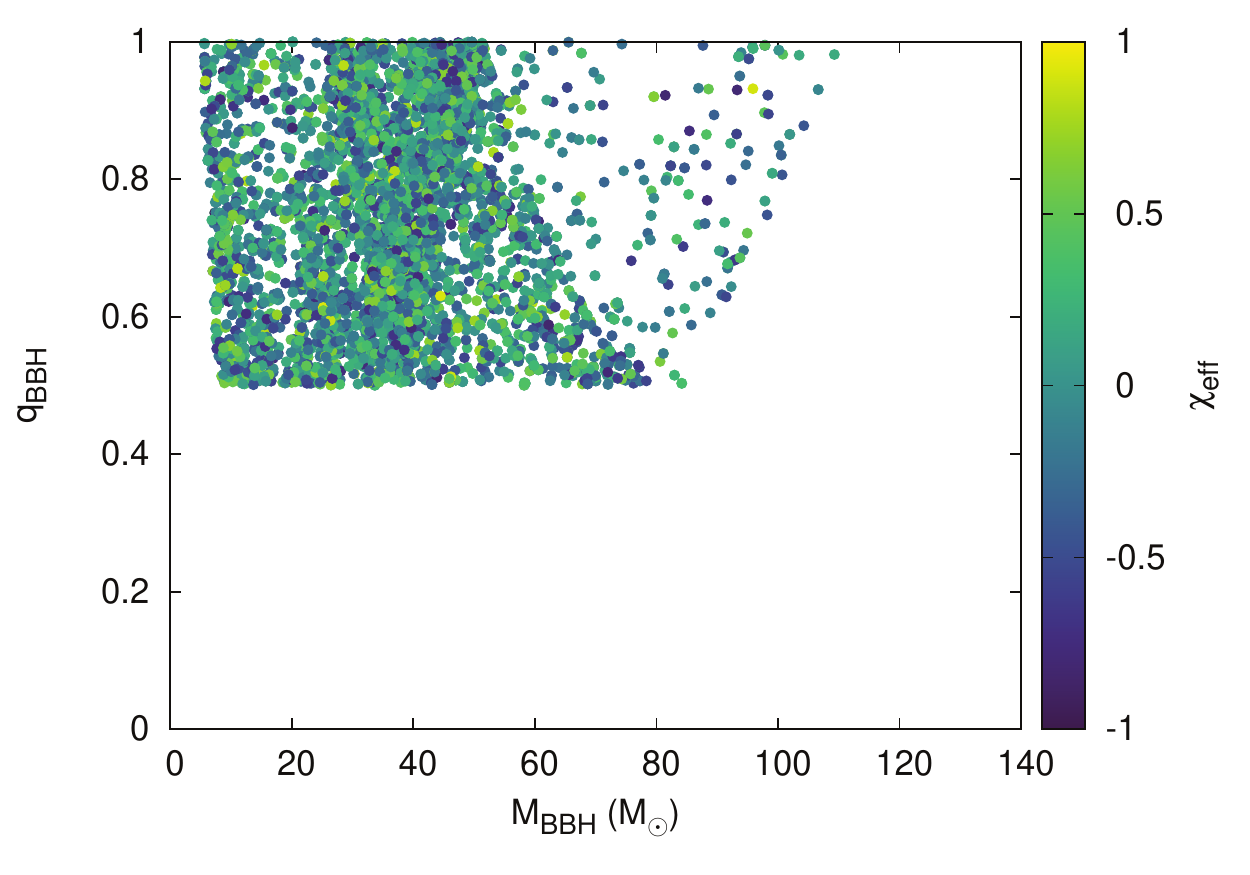}\\
\caption{ Total mass (X-axis) and mass ratio (Y-axis) of the BBH population sample for reference models I0 (top panel), B0 (central panel) and D0 (bottom panel). Color-coding provides information on the BBH components spin amplitude through the {\rm effective spin $\chi_{\rm eff}$}.}
\label{initmodels}
\end{figure}

Figure \ref{initmodels} illustrates the main differences between the three reference models in terms of BBH mass ratio and total mass. We also give a hints on the total BBH spin through the effective spin parameter, defined as
\begin{equation}
\chi_{\rm eff} = \frac{M_1a_1\cos\beta + M_2a_2\cos\gamma}{M_1+M_2},
\end{equation}
where $\beta$ and $\gamma$ are the angles between the orbital angular momentum and the spin of the primary and secondary component, respectively.
Note that isolated binaries evolved with \BSEET lead to progenitor BHs with lower natal spin values, resulting in remnant effective spin values generally lower than for I models, where the natal spins are biased toward high values through Equation \ref{EQanat} in the explored mass range.

\begin{table}
\centering{}
\caption{Main features of our models}
\begin{center}
\begin{tabular}{cccccc}
\hline
\hline
MODEL &  $n_\theta$ & $f_\recy$ & $q_{\rm min}$ & SE & $Z$  \\
\hline\multicolumn{6}{c}{isolated BBHs}\\ \hline
I0        & $\infty$   & $-$           & $-$    & \SSEET  & $2\times 10^{-4}$ \\
I1        & $2$        & $-$           & $-$    & \SSEET  & $2\times 10^{-4}$ \\
I2        & $0$        & $-$           & $-$    & \SSEET  & $2\times 10^{-4}$ \\
I3        & $2$        & $-$           & $-$    & \SSEVN  & $2\times 10^{-4}$ \\
I4       & $2$        & $-$           & $-$    & \SSEET  & $2\times 10^{-2}$ \\
I5       & $2$        & $-$           & $-$    & \SSEVN  & $2\times 10^{-2}$ \\
\hline\multicolumn{6}{c}{isolated BBHs modelled with BSE}\\ \hline
B0        & $\infty$   & $-$      & $-$    & \BSEET  & $2\times 10^{-4}$ \\ 
B1        & $2$   & $-$           & $-$    & \BSEET  & $2\times 10^{-4}$ \\ 
B2        & $0$   & $-$           & $-$    & \BSEET  & $2\times 10^{-4}$ \\ 
B3        & $0$   & $-$           & $-$    & \BSEET  & $2\times 10^{-3}$ \\ 
B4        & $0$   & $-$           & $-$    & \BSEET  & $2\times 10^{-2}$ \\ 
\hline\multicolumn{6}{c}{dynamical BBHs}\\ \hline
D0       & $-$        & $0$           & $0.5$   & \SSEET  & $2\times 10^{-4}$ \\
D1       & $-$        & $0.1$         & $0.5$   & \SSEET  & $2\times 10^{-4}$ \\
D2       & $-$        & $0.1$         & $0.0$   & \SSEET  & $2\times 10^{-4}$ \\
D3       & $-$        & $0.0$         & $0.5$   & \SSEVN  & $2\times 10^{-4}$ \\
D4       & $-$        & $0.0$         & $0.0$   & \SSEVN  & $2\times 10^{-4}$ \\
D5       & $-$        & $0.1$         & $0.5$   & \SSEVN  & $2\times 10^{-4}$ \\
D6       & $-$        & $0.0$         & $0.5$   & \SSEET  & $2\times 10^{-3}$\\
D7       & $-$        & $0.0$         & $0.5$   & \SSEVN  & $2\times 10^{-3}$\\
D8       & $-$        & $0.1$         & $0.5$   & \SSEET  & $2\times 10^{-2}$\\
D9       & $-$        & $0.1$         & $0.5$   & \SSEVN  & $2\times 10^{-2}$\\
\hline
\end{tabular}
\end{center}
\label{tab1}
\end{table}

\subsection{Remnant masses and spins}
\label{remnant}
Numerical relativity allows us to integrate the last phases of the BBH lifetime and obtain reliable information on the gravitational waves produced during the BBH inspiral, plunge, merger and ringdown \citep{pretorius05v,campanelli06,baker06,sperhake2015}.

Since the current signal-to-noise ratio achieved in the LIGO/VIRGO observations does not allow us to clearly disentangle the ringdown phase from the noise, the remnant mass, $M_\rem$, is calculated  is obtained using a handful of fitting formulae calibrated to numerical relativity simulations. 
During the  coalescence, a fraction of the BBH mass is radiated away in form of gravitational energy, thus implying that the remnant mass is only a fraction of the BBH total mass.
 
How this quantity relates to the BBH mass depends on the complex relation connecting the spins of the BBH components, the BBH mass and the mass ratio. Numerical relativity simulations are an unique tool to explore the large phase space and to assess precisely the interplay among all the parameters involved, showing that in general the final BH mass is $\sim 90-100\%$ the BBH total mass \citep{berti07,ajith11,healy14,healy17}. Unfortunately, the complexity and computational cost of these numerical models result in a very sparse coverage of the parameter space. In the last few years, a number of works provided fitting formulae aimed at using simpler expression to describe such complex relations \citep{hemberger13,barausse12,jimenez17}.

\begin{table}
\caption{Known BHs detected through GW emission}
\centering{}
\begin{center}
\begin{tabular}{ccccc}
\hline\hline
name & $M_1$ &$M_2$ &$M_\rem$& $a_\rem$\\
& $\Ms$& $\Ms$ &$\Ms$ &\\
\hline
GW150914  & $36.2^{+5.2}_{-3.8}$ & $29.1^{+3.7}_{-4.4}$ & $62.3^{+3.7}_{-3.1}$ & $ 0.68^{+0.05}_{-0.06}$\\
GW151226  & $14.2^{+8.3}_{-3.7}$ & $ 7.5^{+2.3}_{-2.3}$ & $ 20.8^{+6.1}_{-1.7}$ & $ 0.74^{+0.06}_{-0.06}$ \\
GW170104  & $31.2^{+8.4}_{-6.0}$ & $19.4^{+5.3}_{-5.9}$ & $ 48.7^{+5.7}_{-4.6} $ & $0.64^{+0.09}_{-0.20}$ \\
GW170608  & $12^{+7}_{-2}$ & $ 7^{+2}_{-2}$ & $ 18^{+4.8}_{-0.9}$ & $ 0.69^{+0.04}_{-0.05}$ \\
GW170814  & $30.5^{+5.7}_{-3.0}$ & $ 25.3^{+2.8}_{-4.2}$ & $ 53.2^{+3.2}_{-2.5} $ &$ 0.70^{+ 0.07}_{- 0.05}$ \\
LVT151012 & $23^{+18}_{-6}$ & $ 13.0^{+4}_{-5}$ & $ 35^{+14}_{-4}$&$0.66^{+0.09}_{-0.10}$ \\
\hline
\end{tabular}
\begin{tablenotes}
\item \small{\textbf{Note}: Final spin and mass for GW150914, GW151226 and LVT151012 are taken from \citet{LIGOcat}.}
\end{tablenotes}
\end{center}
\label{Tdata}
\end{table}

Using the available data for the 5 confirmed BBH merger events detected by the LIGO/VIRGO collaboration\footnote{https://losc.ligo.org/events/}, whose main properties are summarized in Table \ref{Tdata},we found that the BH remnant mass is connected to the initial binary mass by a linear relation 
\begin{equation}
M_\rem = \alpha (M_1+M_2), 
\label{sss}
\end{equation}
with $\alpha = 0.957\pm 0.003$.
Although our calculation is based on a simple linear fit to the observed data,  this scaling parameter agrees pretty well with the theoretical expectations for non-spinning, equal mass binaries. More general calculations, which take into account the spins projection along the angular momentum vector \citep{hemberger13,scheel15}, show that the final-to-initial mass ratio can  be reduced down to $\alpha \sim 0.89$ for an equal mass BBH with nearly extremal aligned spin, and increase up to $\alpha=1$ for extreme mass ratio $q<0.1$.  
Although this relation certainly comes out from the General Relativity prescription used to process the observed signal, its simple form allows us to obtain the expected distribution of merger remnants easily.  
However, this lowest order approximation does not take into account the dependence between the remnant BH mass and the progenitor BBH effective spin \citep{hemberger13}, or the mass ratio \citep{healy17,jimenez17}. Hence, in all the models we calculate the remnant mass according to the \cite{jimenez17} fitting procedure, although such approach is tailored to aligned-spin binaries. We note that calculating $M_\rem$ either using Equation \ref{sss} or the \cite{jimenez17} fitting formulae does not severely affect the final distribution.

The remnant spin amplitude is calculated following the fitting formulae provided by \cite{hofmann16} \citep[but see also][]{barausse12, healy14, healy17}, according to which
\begin{eqnarray}
a_\rem =& (1+q)^{-2}[ a_1^2 + a_2^2q^4 + 2 a_1a_2q^2 \cos\alpha + \nonumber \\ 
        & +2\left( a_1\cos\beta + a_2 q^2 \cos\gamma \right) lq + l^2q^2 ]^{1/2},
        \label{finalspin}
\end{eqnarray}
where $\alpha$ is the angle between the BH spins, $\beta$ and $\gamma$ are the angle between the BH spins and the angular momentum, and $l$ is a function of the BBH angular momentum and energy calculated at the last stable circular orbit (see Equation 13 in \cite{hofmann16}).

We refer the reader to \cite{hofmann16} and reference therein for further details about the procedure followed to calculate the merged BH final spin.

In order to validate this procedure, we calculated $a_\rem$ also taking advantage of the fitting formulae provided by \cite{jimenez17} , and making use of the augmentation technique \citep{rezzolla08,hughes03}, which allows us to include the contribute of in-plane spin components in determining the remnant spin.
We found an overall good agreement between the two approaches, with the differences limited to $O(1\%)$ when non-aligned spins are considered.

\section{Results}
\label{results}
\begin{figure}
\centering
\includegraphics[width=8cm]{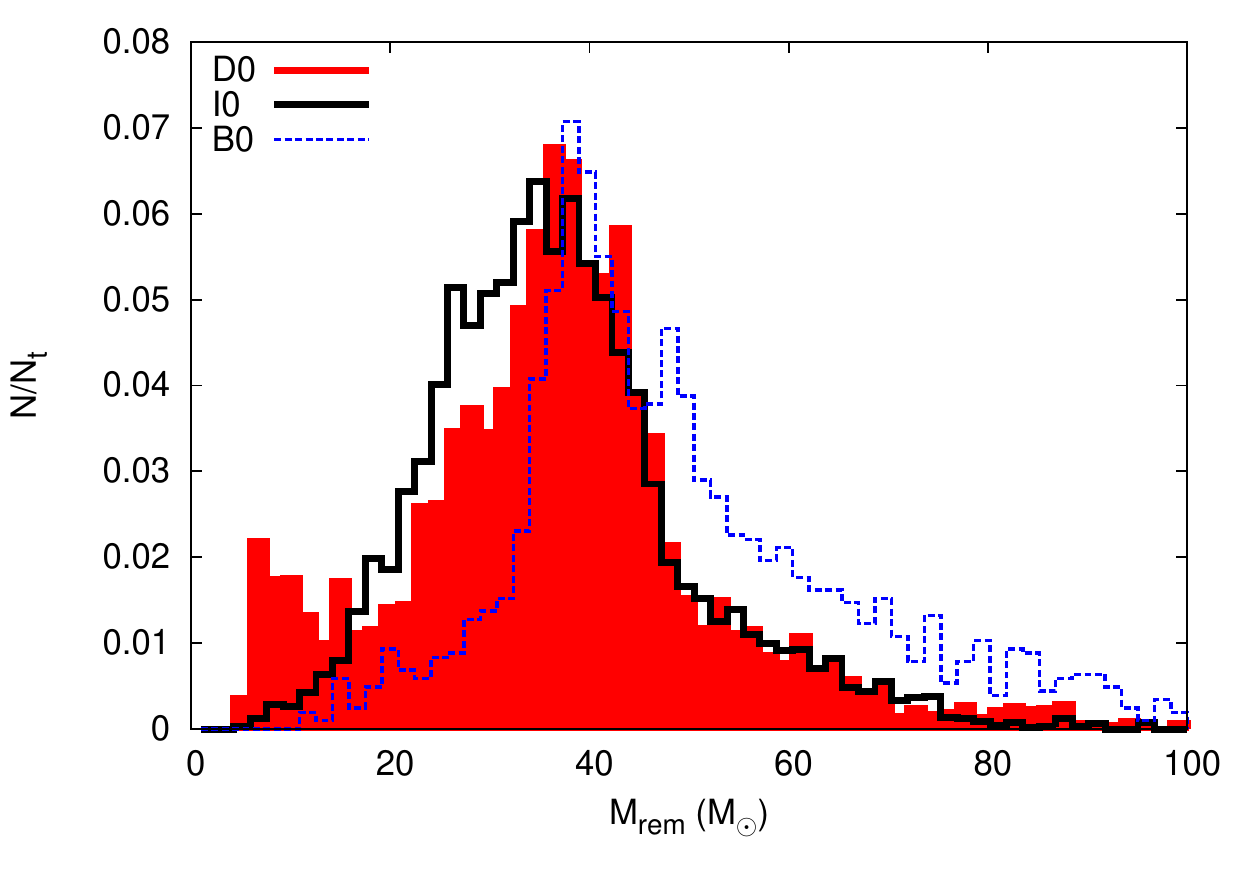}\\
\includegraphics[width=8cm]{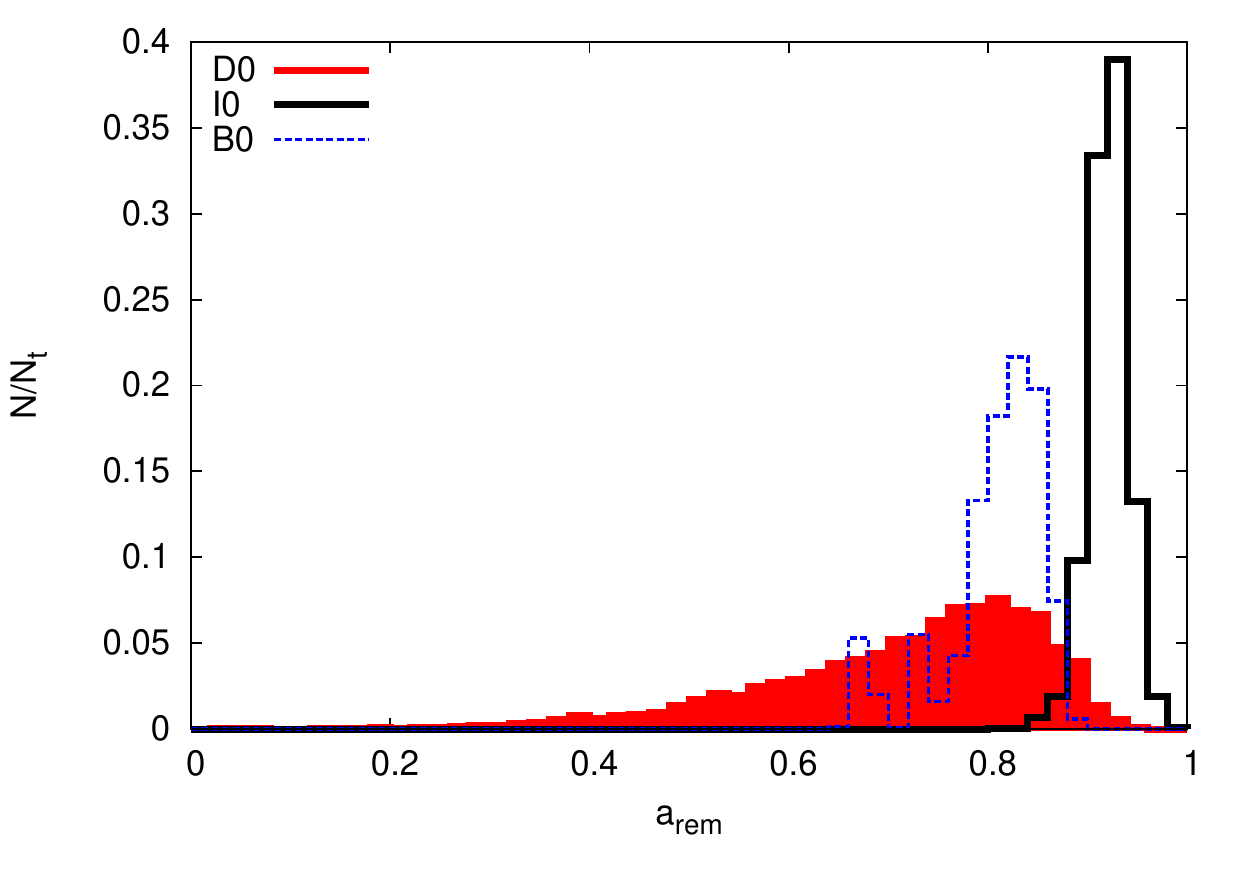}\\
\caption{ Mass (top panel) and spin distribution (bottom panel) for BHs remnants forming in the isolated scenario I (straight black steps), in the isolated scenario B (dashed blue steps) or in the dynamical scenario (red filled boxes).}
\label{histo}
\end{figure}

Figure \ref{histo} shows the mass and spin distributions for remnant BHs formed through the isolated or dynamical scenario.

The top panel of Figure \ref{histo} shows the mass distributions. The mass distribution for dynamically formed BH remnants in reference model D0 has a narrow peak, a longer tail in the BH high-mass end and a greater occurrence of BHs with mass in the range $5-15\Ms$. 

Isolated binaries in the I0 reference model are characterized by a broader mass distribution, with a high-end tail similar to D0 model but with a lower fraction of low-mass BHs. The remnant mass distribution for isolated binaries in the B0 reference model, instead, abruptly drops at BH masses below $\sim 35\Ms$, while it shows an overabundance of BHs with masses above $50\Ms$, compared to I0 and D0 models.

Significant differences are also evident in the remnant spin distribution, shown in the bottom panel of Figure \ref{histo}. In general, it seems quite difficult to generate low-spin BHs. For BH remnants forming through the I0 channel, the spin distribution peaks at values $a_\rem \sim 0.9-0.95$ and is characterized by a very narrow distribution with an evident cut-off below $a_\rem \lesssim 0.8$.

Also BBHs evolved with \BSEET (model B0) are characterized by a very narrow spin distribution, peaking at $a_\rem \sim 0.83$ and ranging mostly between $0.68-0.85$.

On the other hand, the spin of dynamically formed BHs distributes similar to a Gaussian, with a peak at $a_\rem = 0.8$ and a full width half maximum $\sim 0.1$.  Differing from a pure Gaussian, the distribution is characterized by a smooth increase up to the peak value and a sharp decrease beyond that.

The peculiar spin distribution seems to suggest that it is quite hard to produce low-spin BHs from both isolated and dynamical processes. 

The evident peak shifted to relatively large $a_\rem$ values is due to the relation between the progenitor star CO core mass and the BH natal spin.
Therefore, if slowly rotating BHs will be observed in future GW observations runs this could imply the need to revise our knowledge of BHs natal spins.
A simple comparison model can be obtained assuming that the natal spin does not depend on the BH natal mass. We provide an example in Appendix \ref{appB}, assuming the reference model for the dynamical scenario.

The evident differences in the mass and spin distribution represent a potential tool to explore the origin of observed LIGO sources. 

In the following, we will explore how spin misalignment affects the merger products of isolated binaries merger products, and how the final BH mass in dynamical binaries is affected by stellar evolution and repeated mergers. We will relate the distribution of spins and remnant masses described in Figure \ref{histo} to the expected distribution of LIGO/Virgo events by taking into account the selection effects of gravitational radiation observations. The detection volume $V$ for LIGO/Virgo depends upon the masses of the coalescing binary, as higher mass systems generate higher strain amplitudes and are thus observable from greater distances. The exact relationship between mass and distance also depends upon sky location, angle of inclination, and component spins. In addition, the gravitational wave frequency at merger is lower for higher mass systems. This results in a lowering of the signal to noise ratio as the signal spends less time in the sensitivity band of the detector. However, for total masses in the range of $10 \Ms < M_{\rm BBH} < 100 \Ms$, the mass dependence of the detection volume can be well approximated by $V \propto M_{\rm BBH}^{2.2}$ \citep{fishbach17}. Note that this relation is valid at fixed mass ratio, as shown in \cite{fishbach17} Figure 1. 
It must be noted that in their work, \cite{fishbach17} assume that the spin does not affect the BH detectability. Taking into account the components spins can increase the LIGO sensitive volume up to $30\%$, depending on the binary components masses \citep{capano16}, although it is less trivial to show how spins affect the volume dependence on the binary total mass.

Due to the uncertainties related to the effect of mass ratio and spin on the volume - mass relation, we will provide some comparison models assuming $V \propto M_{\rm BBH}^{k}$ with $k=1.5$ and $0$ in Appendix \ref{appC}, to provide the lowest limit possible for our calculations, while leaving to discussion possible effects connected with a steeper relation. 

\subsection{Isolated binaries}

Our isolated binaries in the I models are based on the simplest assumption that the two stars directly evolve to a BH without undergoing any significant mass transfer or common envelope phase. This strongly unrealistic assumption is made so that the I models can serve as a comparison to the self-consistent B models, and to identify the possible signatures of mass transfer and common envelope prescriptions in the merger products. Unlike the B models, the I models do not give the period or eccentricity of the progenitor binary, so we assume that the period of the newly formed BBH is characterized by a flat distribution in the logarithm of the period, limited between $\sim 10^2$~s and $10^4$~days, and that the eccentricity is randomly distributed between 0 and 1. We also looked at a model with the maximum eccentricity limited to $0.2$, but we found that the global distribution of the BBH properties was not significantly affected by this choice.

We also explore the consequences of spin alignment in both the I and B models for isolated binaries. Full alignment of the spins is our simplest assumption. Because the full picture of how the spins evolve within the binary is not clear, we look at the effect that various spin misalignments have on the remnant mass and spin. As described earlier, we varied the parameter $n_\theta$ defined in equation~\ref{ntheta}, choosing $n_\theta = \infty$, 2, or 0.

\begin{figure}
\includegraphics[width=8cm]{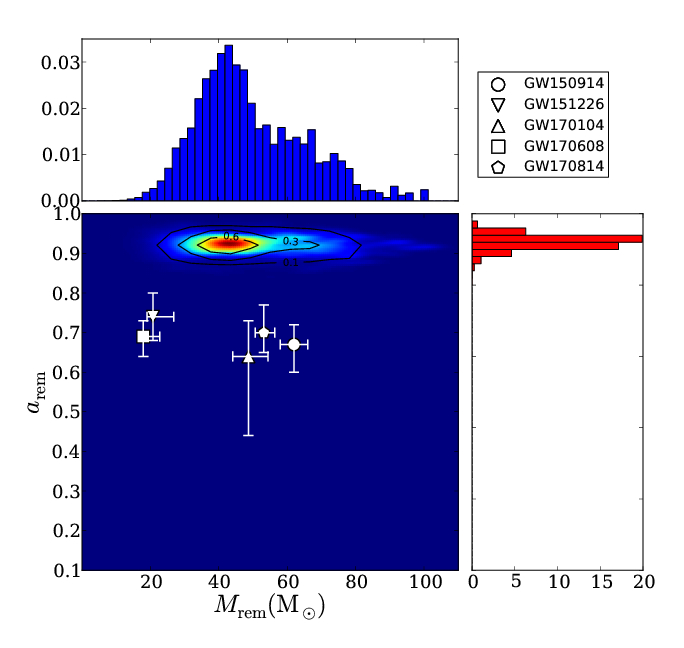}\\
\includegraphics[width=8cm]{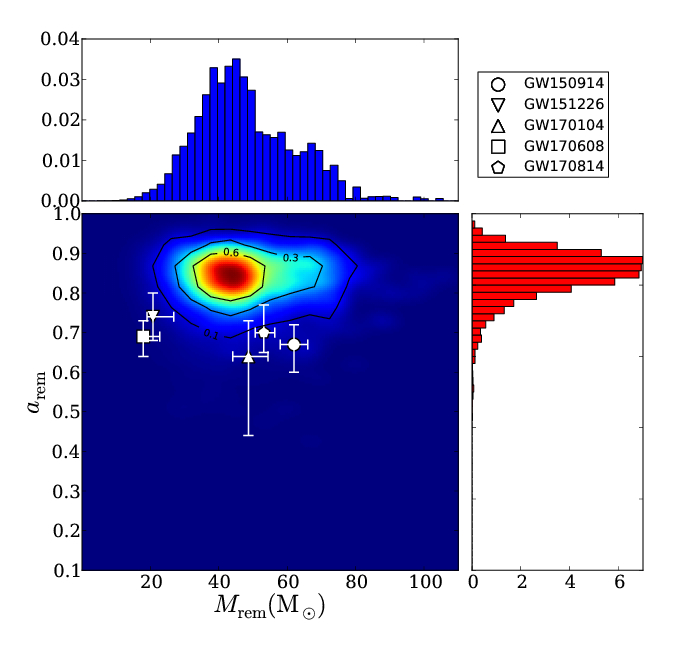}\\
\includegraphics[width=8cm]{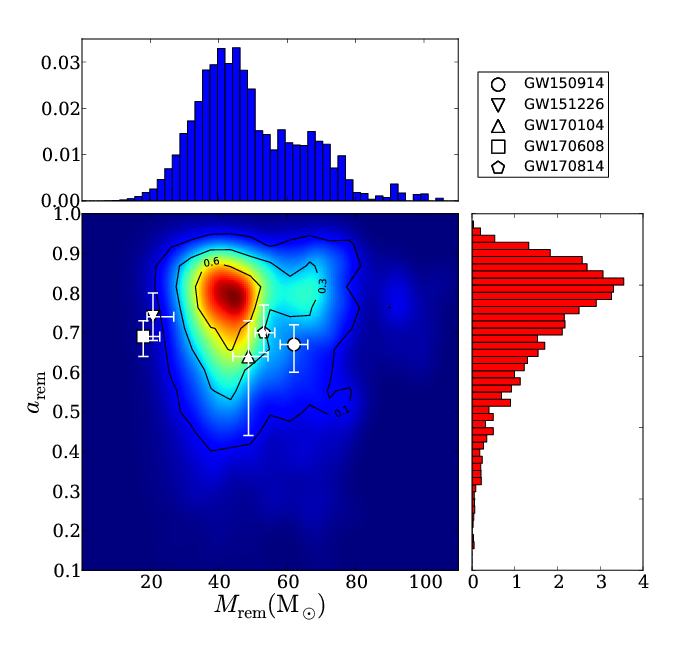}\\
\caption{Final mass and spin of BH remnants formed from field BBHs fully aligned (top panel, model I0), mildly misaligned (central panel, model I1) or completely random spin orientation (bottom panel, model I4).} 
\label{observed1}
\end{figure}

Looking solely at those I models evolved with \SSEET, we see in Figure~\ref{observed1} that when full alignment is considered (model I0) the observed GW sources are quite hard to connect with isolated binary evolution. However, increasing the probability of misalignment brings the theoretical predictions closer to some of the observed sources.
When $n \rightarrow \infty$, i.e.\ full alignment, the spin distribution is strongly peaked at $a_\rem \simeq 0.9$,  
while the maximum BH mass is $M_\rem = 40\Ms$ and the mass distribution is characterized by a smoothly decreasing high-end mass tail. For $n = 2$ (model I1), instead, our results suggest that the spin distribution slightly broadens toward lower spin values,
peaking at $a_\rem \sim 0.85$, while the BH remnant mass distribution remains substantially unaffected.
In the ``chaotic'' case in which the spins of the progenitor BHs are selected randomly (model I4), the $a_\rem$ distribution broadens significantly, smoothly declining up to $a_\rem\sim 0.1$ and peaking at $a_\rem \simeq 0.8$.

Comparing the misaligned models with observations, we find that GW151226 and GW170608 fall outside the predicted distribution. This suggests that either their progenitors have larger metallicities, which imply lower BH natal masses, or that they formed through the dynamical channel. GW170104 and GW170814 fall within the distribution if chaotic spin alignment is assumed, with the lower mass source seeming much more constrained, especially for $n_\theta = 0$. Finally, it seems quite hard to explain the origin of GW150914 with isolated binary evolution, even if a misalignment between spins is taken into account. 

\begin{figure*}
\includegraphics[width=8cm]{isolated_disalign_I1}
\includegraphics[width=8cm]{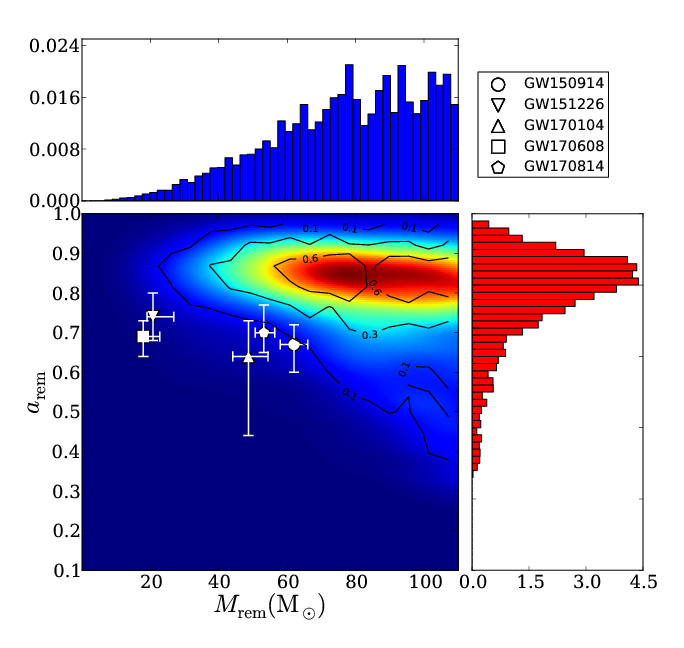}\\
\includegraphics[width=8cm]{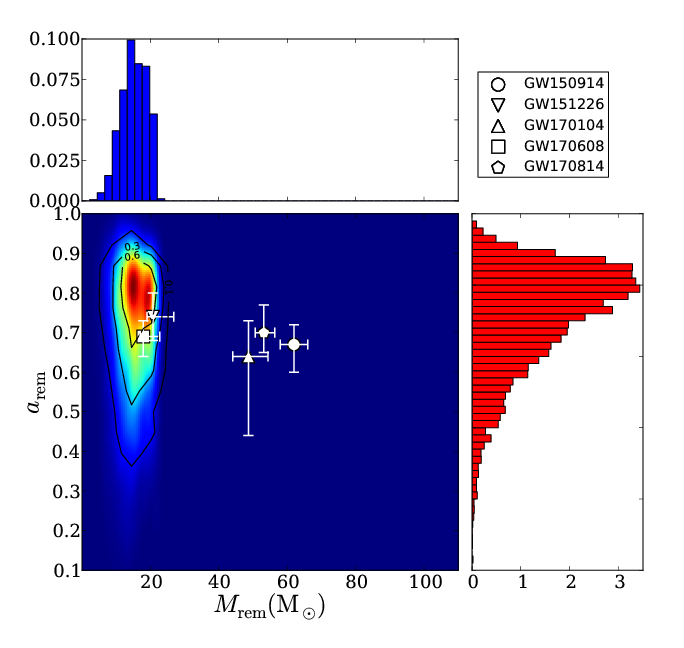}
\includegraphics[width=8cm]{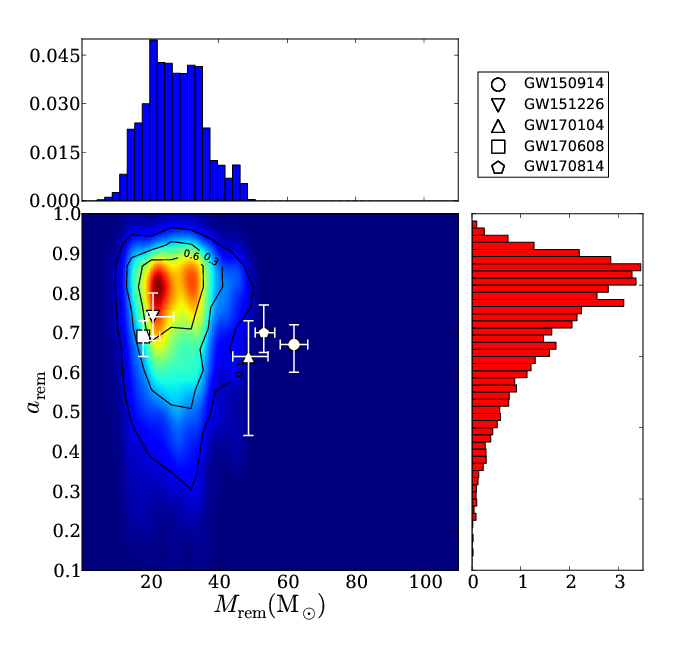}\\
\caption{Remnant mass and spin distribution for misaligned models ($n_\theta = 2$) at low ($Z = 2\times 10^{-4}$, top panels) and for completely misaligned binaries ($n_\theta = 0$) and solar metallicities (bottom panels). We show results obtained with \SSEET (left column panels) and \SSEVN (right column panels). From top to bottom and from left to right panels refer to models I1, I3, I4 and I5.}
\label{observed2}
\end{figure*}

The stellar evolution recipes that are used constitute a crucial ingredient for calculating the BH natal masses and, consequently, the remnant masses. A comparison between results obtained with \SSEET and \SSEVN for a mildly misaligned configuration ($n_\theta = 2$) is shown in Figure~\ref{observed2}. The models shown in the top panels are characterized by low-metallicity ($Z = 2\times 10^{-4}$), while for the bottom panels we assumed solar values ($Z = 0.02$).

Comparing the left and right panels in Figure~\ref{observed2} clearly shows the differences between the two stellar evolution codes. \SSEVN produces much more massive BHs---especially at low metallicities, where it predicts remnant BH masses peaking at values $\gtrsim 90 - 100\Ms$. The overlap with observations in this case can be achieved, in principle, by assuming metallicities larger than $Z=2\times 10^{-4}$ and a higher degree of misalignment.

Large BH natal masses significantly affect the $M_{\rm rem} - a_{\rm rem}$ distribution for misaligned isolated binaries with low metallicity, as shown in the right column panels of Figure~\ref{observed2}. In this case, observable remnant masses most likely fall in the range $60 - 100\Ms$, while their spin lies mostly in the range of 0.7 and 1.

At solar metallicity, both approaches explain quite well the origin of GW151226 and GW170608, provided we decrease the misalignment parameter $n_\theta$. In the case of \SSEET, we note that a better overlap between the model and observations would be achieved at slightly sub-solar values of $Z$, as opposed to \SSEVN, for which super-solar metallicities are required to move the $M_{\rm rem}$ peak to smaller values.

\subsection{Isolated binaries evolved with BSE}

The simple procedure described above does not take into account the evolutionary phases of the binary prior to the formation of the coalescing BH-BH. 
On another hand, the choice of a specific package for following binary star evolution is made more difficult by the fact that there are currently many packages providing different results for the same initial conditions. As discussed above, for the sake of coherence we used \BSEET, an updated version of the \texttt{BSE} package currently implemented in the \nbodysix N-body code.

Following the set-up described in the previous section, we paired the progenitor systems in such a way that the primary component is selected according to a power-law distribution with slope $-2.3$ \citep{belczynski16b} and the secondary is kept randomly in the mass range $18-150 \Ms$, in order to focus only on systems that may potentially evolve to form heavy BH-BH pairs.

To model the BBH progenitor binaries, we assigned an initial period distributed logarithmically $\Gamma(\log P) \propto (\log P)^{1/2}$, limited between $P_{\rm min} = 2$ days and $P_{\rm max} = 3000$ days, while the initial eccentricity is drawn according to a $P(e) \propto e^{-0.42}$ distribution \citep{sana12,belczynski17}. Note that constraining the orbital period implies a constraint on the initial semi-major axis, since $a^3 \propto P^2$.
Each binary is then passed to \BSEET and evolved to a maximum time of 13 Gyr. 
Among the whole sample, we select only those binaries evolving into a BBH and merging within 13 Gyr. Following this strategy we found that $24\%$ of the modelled BBHs undergo a merger within this time-frame.

We then use the component masses and spins, which are provided directly from the \BSEET output, and calculate the corresponding remnant mass and spin.

\begin{figure}
\includegraphics[width=8cm]{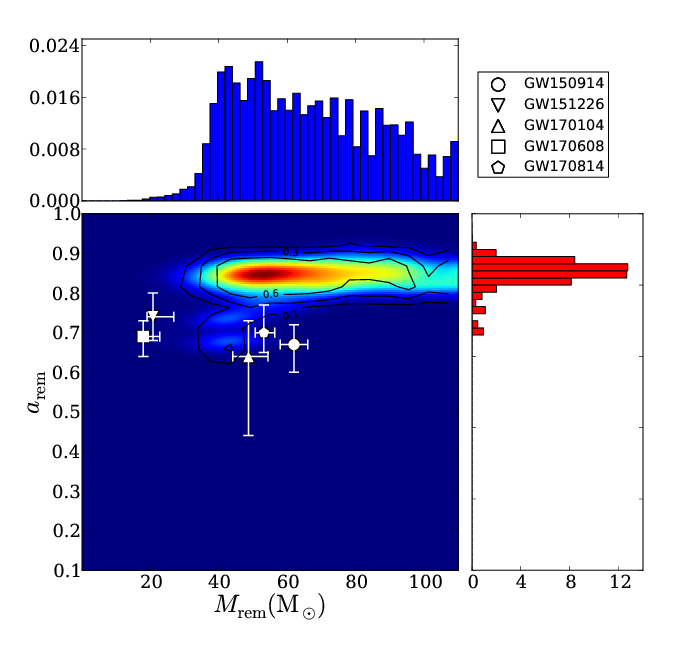}\\
\includegraphics[width=8cm]{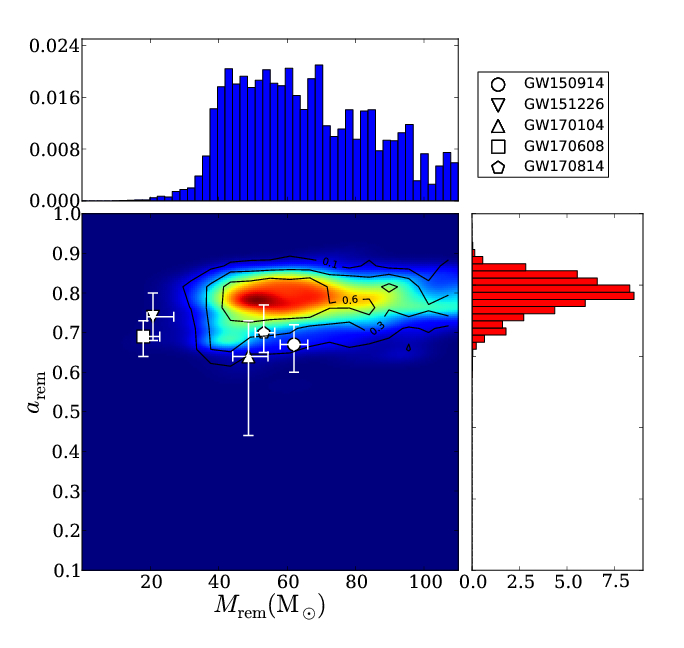}\\
\includegraphics[width=8cm]{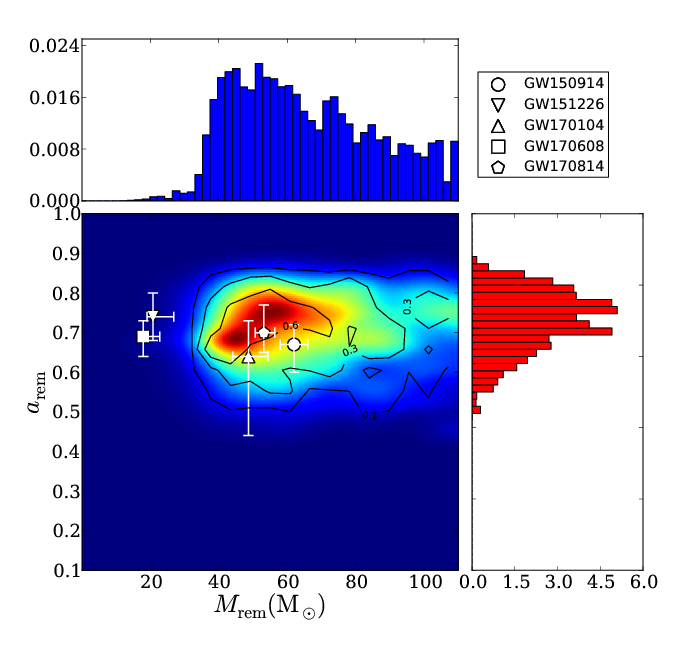}\\
\caption{Remnant mass and spin distribution for models evolved self-consistently with BSE, assuming fully aligned spins (top panel, model B0), mildly misaligned (central panel, model B1) or completely random spin orientation (bottom panel, model B2).}
\label{obsBSE1}
\end{figure}

Figure \ref{obsBSE1} shows a comparison between B models in which we assumed full spin alignment (model B0), mild alignment (B1) or chaotic spin distribution (B2) similar to the I models described in the previous section.

The results differ significantly from simple I models. The $M_\rem$ distribution is characterized by a large fraction of objects with masses above $80\Ms$, while the spin distribution is shifted to lower values, peaking at around $a_\rem \simeq 0.8$ and being narrower compared to the I models.

\begin{figure}
\includegraphics[width=8cm]{isolated_BSE_disalign_0_B2}\\
\includegraphics[width=8cm]{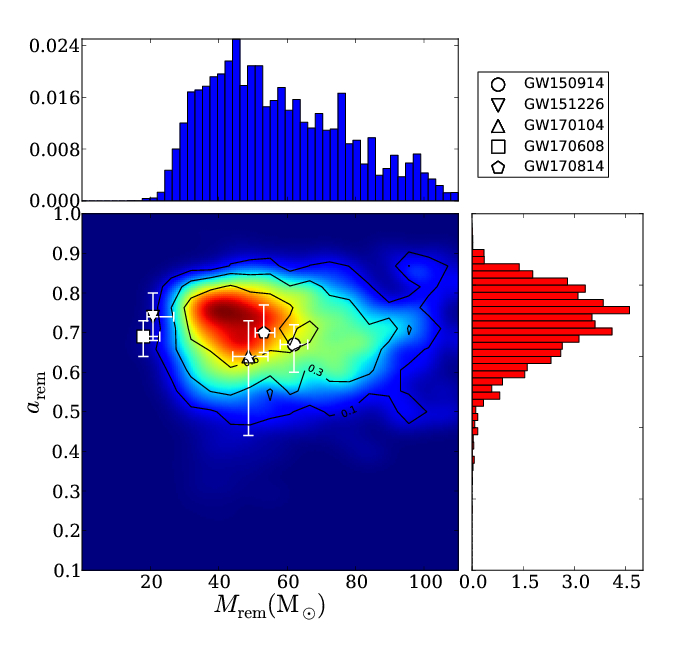}\\
\caption{Remnant mass and spin distribution for models evolved self-consistently with BSE, assuming completely random spin orientation and metallicity $Z = 2\times 10^{-4}$ (top panel, model B2) or $Z = 2\times 10^{-3}$ (bottom panel, model B3).}
\label{obsBSE2}
\end{figure}

Interestingly, assuming full or mild alignment (models B0 and B1) only marginally affects the $M_\rem - a_\rem$ distribution, while a random distribution of the BH spins significantly changes the global distribution. 
Indeed, in model B2 the $a_\rem$ distribution is broader and peaks at lower values $\sim 0.7$, while $M_\rem$ seems flatter, extending beyond $100 \Ms$. All three heavy LIGO sources (GW150914, GW170104 and GW170814) fall within the $M_\rem - a_\rem$ distribution of model B2, while partial alignment seems to able to explain only the GW170814 origin.
For the case in which BH natal spins are randomly oriented, the remnant spin distribution does not change much at sub-solar metallicity values, $Z = 2\times 10^{-3}$ (model B3), while the lower BH natal mass due to metal-dependent stellar winds translate into a $M_\rem$ distribution wider than in model B1 peaking at $M_\rem \simeq 40 \Ms$, as shown in Figure \ref{obsBSE2}. 

Shifting to solar metallicity (model B4) drastically reduces the number of merging BBHs, leading to a small $0.03\%$ merging probability. On another hand, intriguingly we found a high number of merging events involving a BH and a MS star, having this process an occurrence of $~ 3.3\%$.

This has profound implications on the interpretation of the origin of sources GW151226 and GW170608. Indeed, if they have formed via classical isolated binary evolution, they should have formed in an extremely rich stellar population characterized by a high metal content, in order to achieve a reliable formation probability. On another hand, it is also possible that the stellar evolution processes included in the current version of \BSEET need to be improved.
A further, simpler, possibility is that these ``low-mass'' BHs formed via dynamical interactions in a metal-rich cluster like an open or a young massive cluster, as we discuss in the next section.

\subsection{Dynamical binaries}

In section~\ref{pair}, we described the parameters that are likely to affect the $M_{\rm rem}-a_{\rm rem}$ distribution. These are the BBH mass ratio, total mass, metallicity, and fraction of recycled BHs. Our reference model, D0, uses a BBH mass ratio $q>q_{\rm min} = 0.5$, a recycled fraction of $f_{\rm recy} = 0$, and a low metallicity with $Z = 2\times 10^{-4}$. The masses of the BHs are computed using \SSEET.

Our simplest comparison model, D1, assumes that 10\% of the merged BBHs are retained within the parent cluster and can undergo a later meger. Thus, we have $f_{\rm recy} = 0.1$.

Our choice of $q_{\rm min}$ was based on the assumption that the pairing of BHs in dense stellar environments is primarily drived by mass segregation, so that the more massive BHs in the cluster are more likely to pair up. In model D2, we relax this assumption and let $q_{\rm min}=0$, while retaining $f_{\rm recy}=0.1$.

\begin{figure}
\includegraphics[width=7cm]{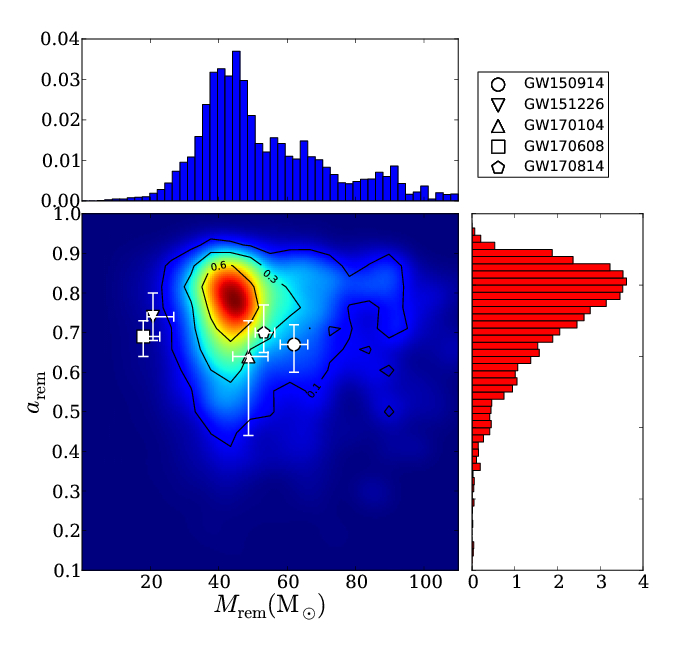}\\
\includegraphics[width=7cm]{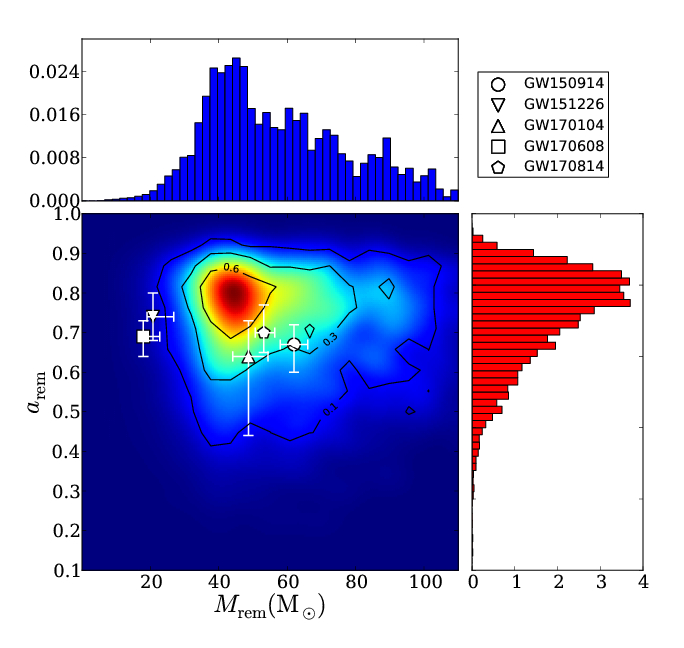}\\
\includegraphics[width=7cm]{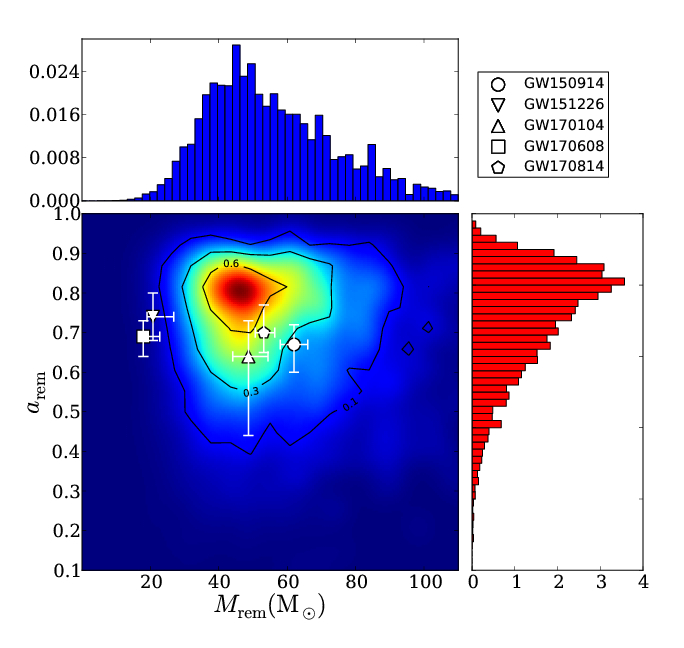}\\
\caption{Remnants mass and spin for dynamically formed BBHs assuming $f_\recy=0$ (top panel, model D0), $f_\recy = 0.1$ (central panel, model D1), or that both $f_\recy = 0.1$ and $q_{\rm min} = 0$ (model D2).}
\label{observed3}
\end{figure}

Figure~\ref{observed3} shows the final BH mass and spin distributions for the three models, D0, D1, and D2. The spin distribution is quite similar over all three models, characterized by a single peak at $a_{\rm rem} =0.8$ with a broad tail to lower spins and a sharp drop-off toward maximally spinning black holes. In model D0, the mass distribution is peaked around $M_{\rm rem}\sim 40\Ms$ with a shallow tail to higher masses. When we allow for a small fraction of recycled mergers and relax the mass ratio constraint, we see an increase in the distribution at higher masses, extending out to $M_{\rm rem}\sim 60\Ms$ with a larger full half maximum width of $\delta M_{\rm rem} \sim 30\Ms$.

Comparing the models with observations suggests that the assumption of random BH pairing leads, on average, to a better constraint on the origin of the heaviest detected sources, namely GW150914 and GW170814. As in the case of isolated binaries, the low mass sources (GW170608 and GW151226) are harder to explain with low metallicity progenitors. These sources are more likely to have formed from sub-solar or solar metallicity progenitors. The intermediate mass source GW170104 is well constrained by this scenario, even without taking into account any restrictions on the mass ratio or the fraction of recycled mergers.

In the extremely dense nuclear star clusters sitting in the centres of many $10^7-10^{11}\Ms$ galaxies, an even larger recycling fraction is possible. Our results indicate that GW150914 may have originated in such an environment.

\begin{figure}
\includegraphics[width=7cm]{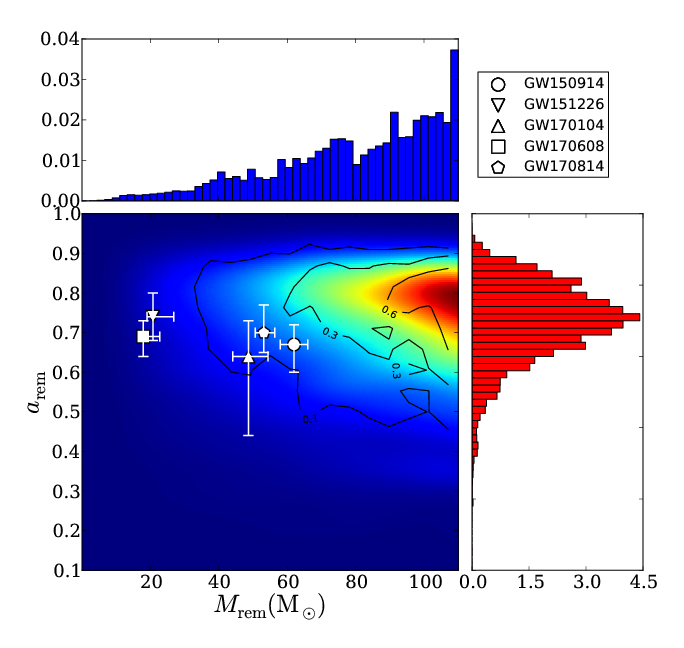}\\
\includegraphics[width=7cm]{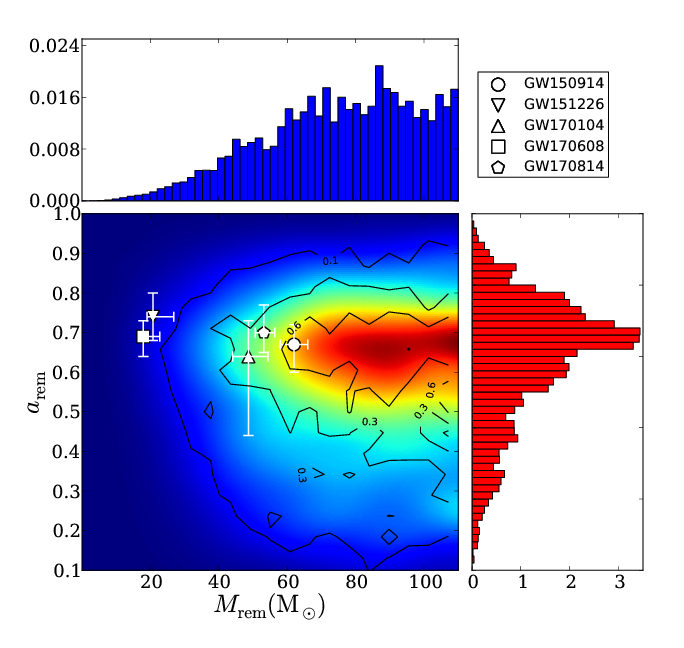}\\
\includegraphics[width=7cm]{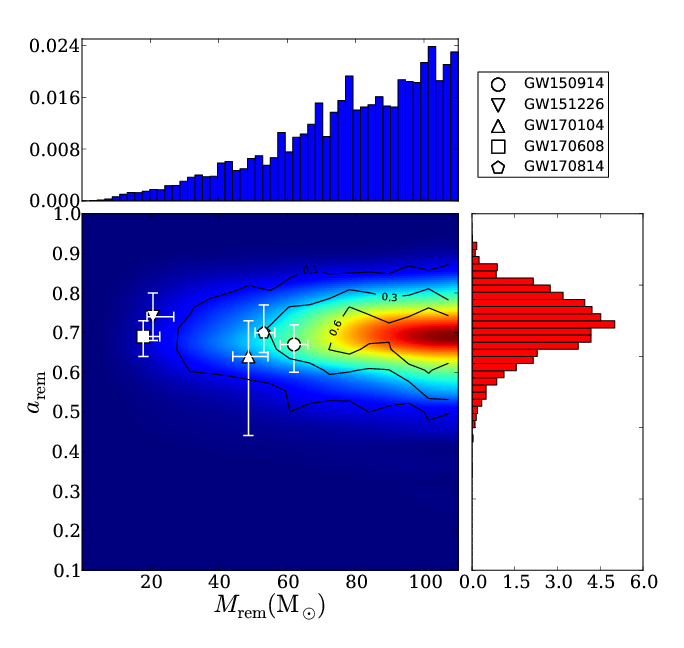}\\
\caption{Remnant mass and spin distribution calculated with \SSEVN formulae assuming $Z = 2\times 10^{-4}$. Top panel: we assumed $q_{\rm min} = 0.5,~ f_\recy = 0$ (model D3). Central panel: $q_{\rm min} = 0,~ f_\recy = 0$ (model D4). Bottom panel: $q_{\rm min} = 0.5,~ f_\recy = 0.1$ (model D5).}
\label{observed4}
\end{figure}

When we introduce the \SSEVN stellar evolution recipes, the picture changes dramatically, as shown in Figure~\ref{observed4}. The peak of the mass distribution moves to much larger values with $M_{\rm rem} \gtrsim 90-100\Ms$. The significant shift to higher masses is primarily due to the weighting toward higher detection probability for high-mass systems. This tends to amplify the influence of the highest mass systems.

In this case, dropping the constraint on the minimum value of the BBH mass ratio, i.e. $q_{\rm min} = 0$, leads the mass distribution to broaden,  while the spin distribution slightly widens over the $0.1-0.9$ values range, as shown in the central panel of Figure \ref{observed4}. Finally, assuming a small recycling fraction and $q_{\rm min} = 0.5$ (the bottom panel in Figure \ref{observed4}) leads the peak to shift toward even more large masses. 

\begin{figure}
\includegraphics[width=8cm]{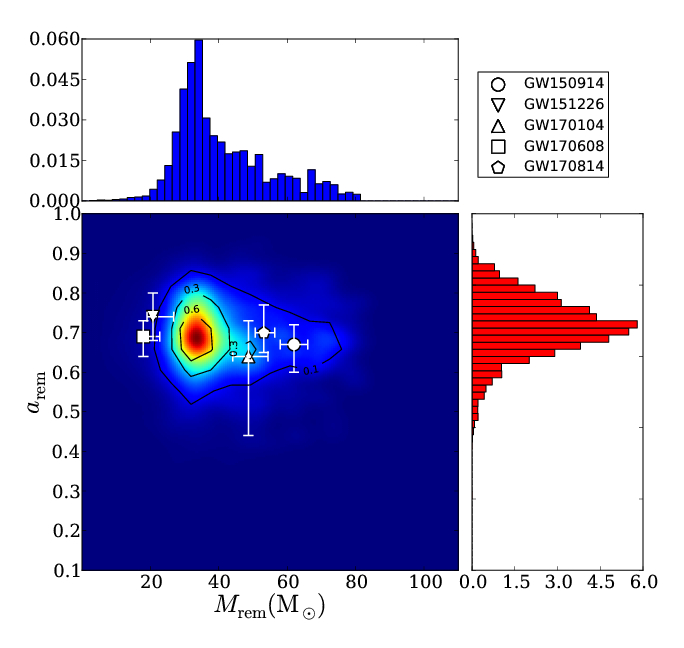}\\
\includegraphics[width=8cm]{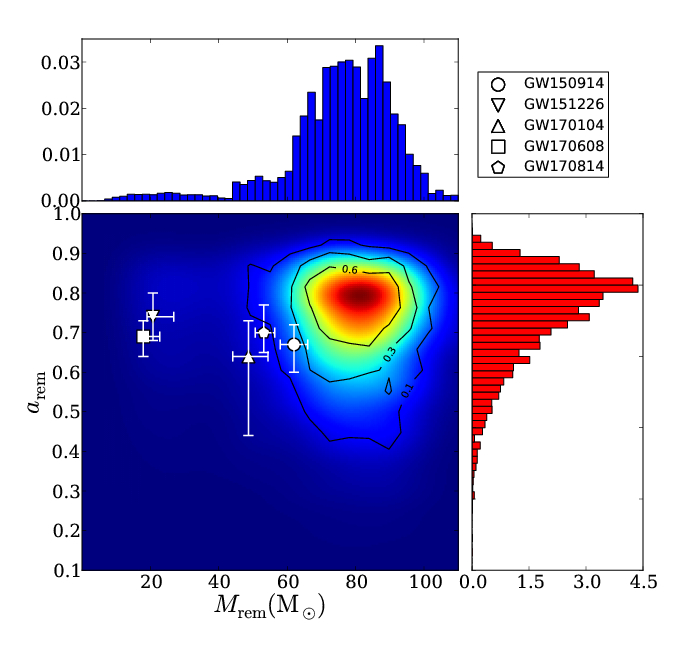}\\
\caption{As in Figure \ref{observed4} but here we assumed $Z = 2\times 10^{-3}$, $q_{\rm min} = 0.5$ and $f_\recy = 0.1$. Top panel refers to \SSEET stellar evolution (model D6) while bottom panel refers to \SSEVN results (model D7).}
\label{observed5}
\end{figure}

If it turns out that the \SSEVN stellar evolution recipes are the most reliable for BH formation, then these results suggest that none of the observed GW sources formed in a low metallicity environment. Only GW150914 falls marginally within the low metal distributions. Moving toward higher metallicities can alleviate this tension. For example, Figure~\ref{observed5} shows how the remnant mass and spin distributions change if we assume a higher metallicity of $Z=2\times 10^{-3}$, $q_{\rm min}=0.5$, and $f_\recy = 0$ for both stellar evolution recipies, \SSEET (top panel) and \SSEVN (bottom panel). Note, however, that the \SSEVN prediction is still too large to explain the origin of 4 out of the 5 detected GW events. At the same time, the \SSEET results suggest that the LIGO sources that are heavier than $40\Ms$ likely formed in extremely low-metal environments.

\begin{figure}
\includegraphics[width=8cm]{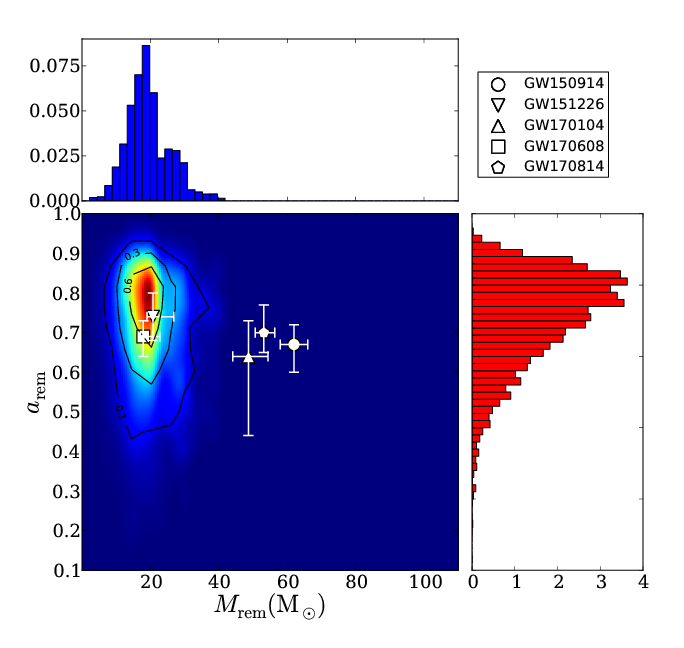}\\
\includegraphics[width=8cm]{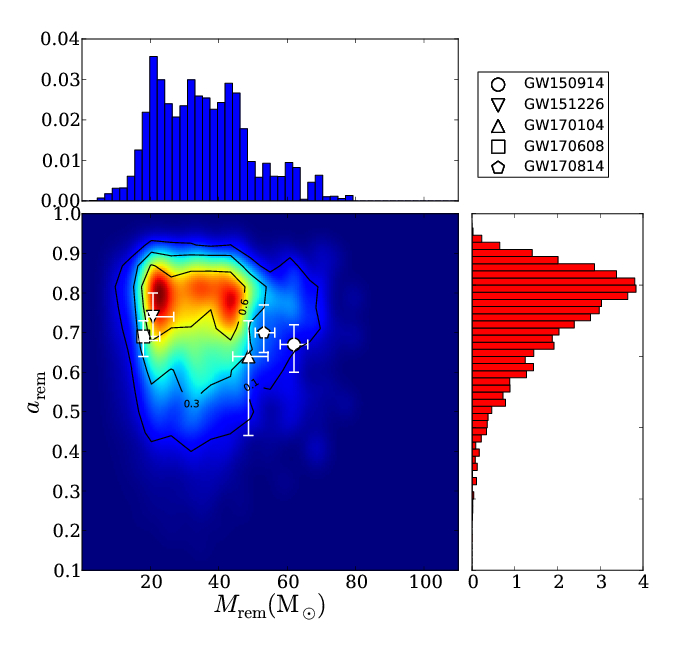}\\
\caption{As in Figure \ref{observed4} but here we assumed $Z = 2\times 10^{-2}$, $q_{\rm min} = 0.5$ and $f_\recy = 0.1$, for different stellar evolution recipes: \SSEET (top panel, model D8) or \SSEVN (bottom panel, model D9).}
\label{observed6}
\end{figure}

In order to investigate the possibility that LIGO sources originated 
in a solar-metallicity environment we provide two more models, shown in Figure \ref{observed6}. In this case, we assumed $q_{\rm min}=0.5$ and a small recycling fraction, $f_\recy = 0.1$. If we rely on \SSEET recipes, as shown in the top panel of Figure \ref{observed6}, our results suggest that only low mass BHs (GW151226 and GW170608) likely formed from metal rich stars. On another hand, \SSEVN can also explain GW170104, while the remaining sources fall outside the distribution. 

\section{Discussion}

In this section we discuss event rates inferred from our models and connect them with an estimte of the number of detections needed to distinguish between formation channels. Finally, we conclude with a more speculative discussion of the likely formation scenarios for each of the BBH mergers observed by LIGO.

\subsection{Merger rates}

In our low-metallicity B models, we found that roughly 30\% of all simulated BBH would merge within a Hubble time. We can use this information to infer the BBH merger rate in the local universe due to isolated binaries. (Note that we do not consider the I models here.)

We begin by assigning a formation time for each merging BBH that has been randomly chosen between 0 and 10 Gyr. We then use the time between formation and merger as determined by the simulation to define a delay time, which is the sume of the formation time and the time to merge. This gives us the time of merger for the BBH. In order to be observable by LIGO, a BBH must merger on a timescale comparable to the age of its host galaxy. Assuming that the age of a typical Milky-Way type host galaxy is $T_{\rm H} \sim 9.8~{\rm Gyr}$, we include only those systems with delay times between 9 and 13 Gyr~\citep{dominik12}. With this information, we define the merger rate to be
\begin{equation}
\Gamma_{\rm iso} = \mathcal{F} f_{\rm IMF} \xi_{\rm IMF} f_{\rm bin} N_{\rm *,MW} N_{\rm G,MW} V_z^{-1} T_{\rm H}^{-1}\,,
\label{mrate}
\end{equation}
where $\mathcal{F} = 8\times 10^{-3}$ is the fraction of BBH with delay times less than $T_{\rm H}$. Because we are only simulating stars with masses above $18\Ms$, we select a fraction $f_{\rm IMF}=10^{-3}$ of stars from the IMF. The factor $\xi_{\rm IMF}$ represents the probability that a BH progenitor will have another BH progenitor as its companion. Since we assumed a flat mass distribution for the companion stars, we find $\xi_{\rm IMF} = 0.88$. The binary fraction is taken to be $f_{\rm bin} = 0.5$ \citep[see for instance][]{sana13,giacobbo18}, while the number of stars in a typical Milky-Way type galaxy is assumed to be $N_{\rm *,MW} = 10^{11}$. We assume the LIGO observable horizon to be $D_{\rm hor}=460~{\rm Mpc}$, corresponding to a redshift of $z \sim 0.1$ and a comoving volume of $V_z = 0.313~{\rm Gpc}^3$. The total number of Milky-Way type galaxies within this volume is found from~\citet{abadie10} to be
\begin{equation}
N_G = 0.0116 \frac{4\pi}{3}\left(\frac{D_{\rm hor}}{{\rm Mpc}}\right)^3 \left(2.26\right)^{-3}, 
\end{equation}

Plugging all these numbers into Equation~\ref{mrate} leads to a BBH merger rate for the isolated channel in the low-$Z$ regime of $\Gamma_{\rm iso} = 46~{\rm yr}^{-1}~{\rm Gpc}^{-3}$

For the dynamically formed binaries, we define the merger rate to be
\begin{equation}
\Gamma_{\rm dyn} = K_{\rm mer} N_{\rm G,MW} V_z^{-1} T_{\rm H}^{-1},
\label{drate}
\end{equation}
where $N_{\rm G,MW}$, $V_z$, and $T_{\rm H}$ are as before, and $K_{\rm mer}$ is a parameter that serves to quantify the typical number of mergers expected for a Milky-Way type galaxy. We estimate $K_{\rm mer}$ using $K_{\rm mer} = \langle N_{\rm mer}\rangle N_{\rm GC}$, where $N_{\rm mer}$ is the total number of mergers per massive star cluster (GC) and $N_{\rm GC}$ is the total number of GC per Milky-Way type galaxy. Following \citet{askar17}, the total number of BBH mergers per GC is linked to the initial mass of the GC via a simple power law, $N_{\rm mer} = \alpha M_{\rm GC}^\beta$, with the coefficients $\alpha$ and $\beta$ depending on the initial cluster metallicity. If we also assume that the GC initial mass distribution follows a simple power law, $f\left(M_{\rm GC}\right) \propto M_{\rm GC}^{-2}$, then the GC average mass is
\begin{equation}
\langle M_{\rm GC}\rangle = - \frac{\log\left(M_{\rm max}/M_{\rm min}\right)}{M_{\rm max}^{-1}-M_{\rm min}^{-1}} \simeq 2.7\times 10^5\Ms
\end{equation}
where $M_{\rm max} = 10^7\Ms$ and $M_{\rm min} = 5\times 10^4\Ms$ are the GC mass function limiting values \citep{gnedin14,webb15}. We assume that 1\% of the total galaxy mass $M_g$ is made up of GC, $M_{\rm GC}=0.01M_g$ \citep{ASCD14,gnedin14,belczynski18}. Thus, the total number of GC per Milky-Way type galaxy is given by $N_{\rm GC} = 0.01M_g/\langle{M_{\rm GC}}\rangle$ \citep{ASCD14,belczynski18}. Thus we have
\begin{equation}
K_{\rm mer} = \langle N_{\rm mer}\rangle N_{\rm GC} \equiv \alpha\langle M_{\rm GC}\rangle^\beta\times\frac{0.01M_g}{\langle M_{\rm GC}\rangle}\,.
\end{equation}
Substituting in the relevant quantities in Equation~\ref{drate} leads to $\Gamma_{\rm dyn} = 7.5~{\rm yr}^{-1}~{\rm Gpc}^{-3}$. Note that although the procedure described above is quite simple, the resulting merger rates agree pretty well with previous estimates \citep[see for instance][]{askar17,rodriguez16}.

\subsection{How many detections are needed to spot differences in different formation channels?}
As we have shown in the previous sections, different scenarios can lead to different distributions in the BH final mass and spin
plane. In this subsection, we use our models to create a sample of mock observations, to compare with LIGO current and future detections. 

Our simplest approach is to gather the results obtained from models D0, i.e. dynamically formed binaries with $q_{\rm min} = 0.5$ and $f_{\rm recy}=0$, and B2, i.e. isolated binaries modelled with \BSEET assuming complete misalignment $n_\theta = 0$. Note that we limit our analysis only to models with $Z = 2\times 10^{-4}$.

We extract a BBH from the mixed sample comprised of D0-B2 binaries, assuming that ``isolated mergers'' are 
more frequent than ``dynamical mergers'', as borne out by our rate estimates of the previous subsection. In fact, according to the recent literature, the BBH merger rate from the isolated channel is expected to be $\sim 2-5$ times larger than for the dynamical channel, being $\sim 10-500$ yr$^{-1}$ Gpc$^{-3}$ for isolated binaries \cite{belczynski17,marchant16,mandel16,mapelli17}, $\sim 1-50$ yr$^{-1}$ Gpc$^{-3}$ for young massive and old globular clusters \citep{banerjee16,banerjee17,askar17,rodriguez15,rodriguez17,samsing17}, and of the order of the unity for galactic nuclei \citep{antonini16b,ASG18,ASCD17b,bartos17,hoang18}. 
Following the literature, our selection is made in such a way that there is a $66.66\%$ chance to select an isolated BBH and $33.33\%$ to select a dynamical BBH.

For the sake of coherence, we also assume that the probability of extracting a candidate scales with the binary mass as $M_{\rm BBH}^{2.2}$, in order to account for the volume-dependent LIGO sensitivity  \citep{fishbach17}.

\begin{figure*}
\centering
\includegraphics[width=8cm]{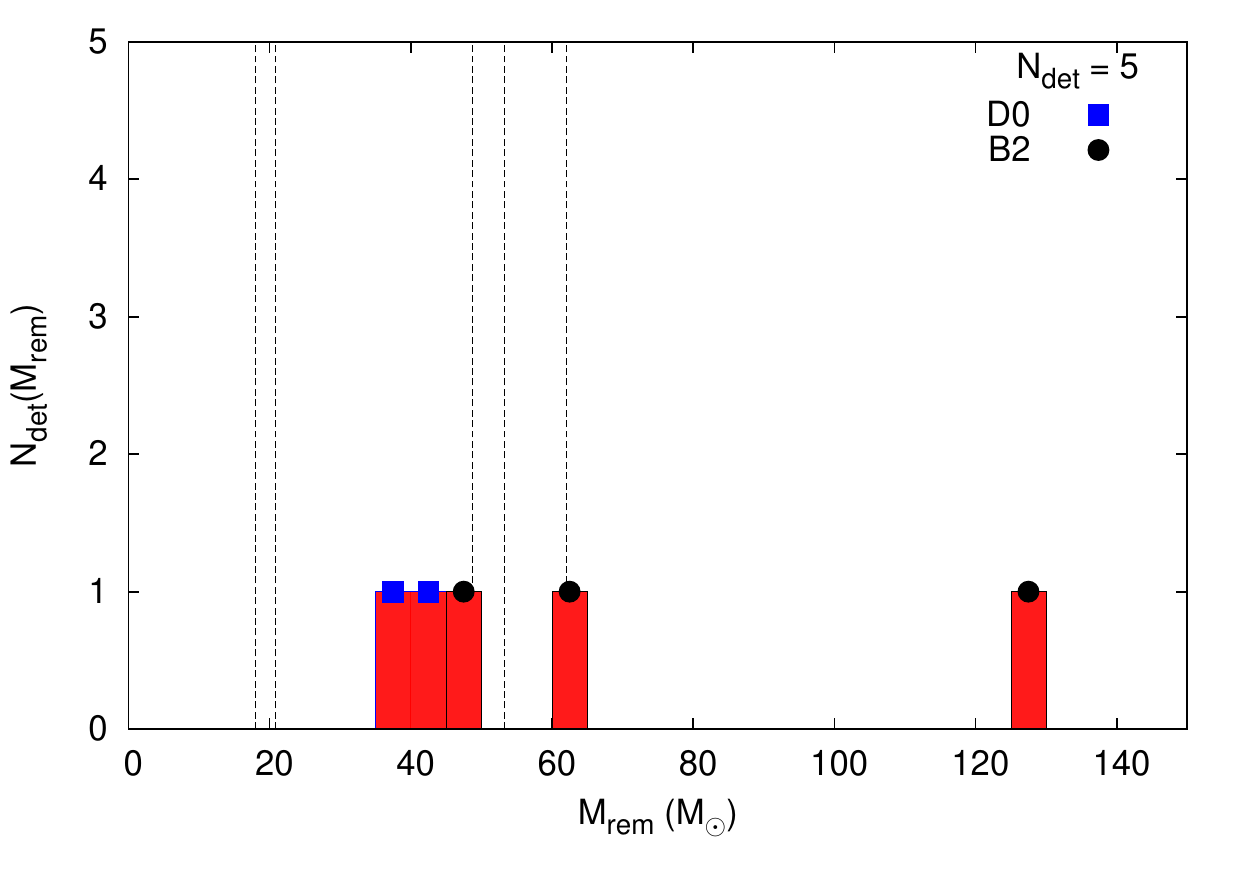}
\includegraphics[width=8cm]{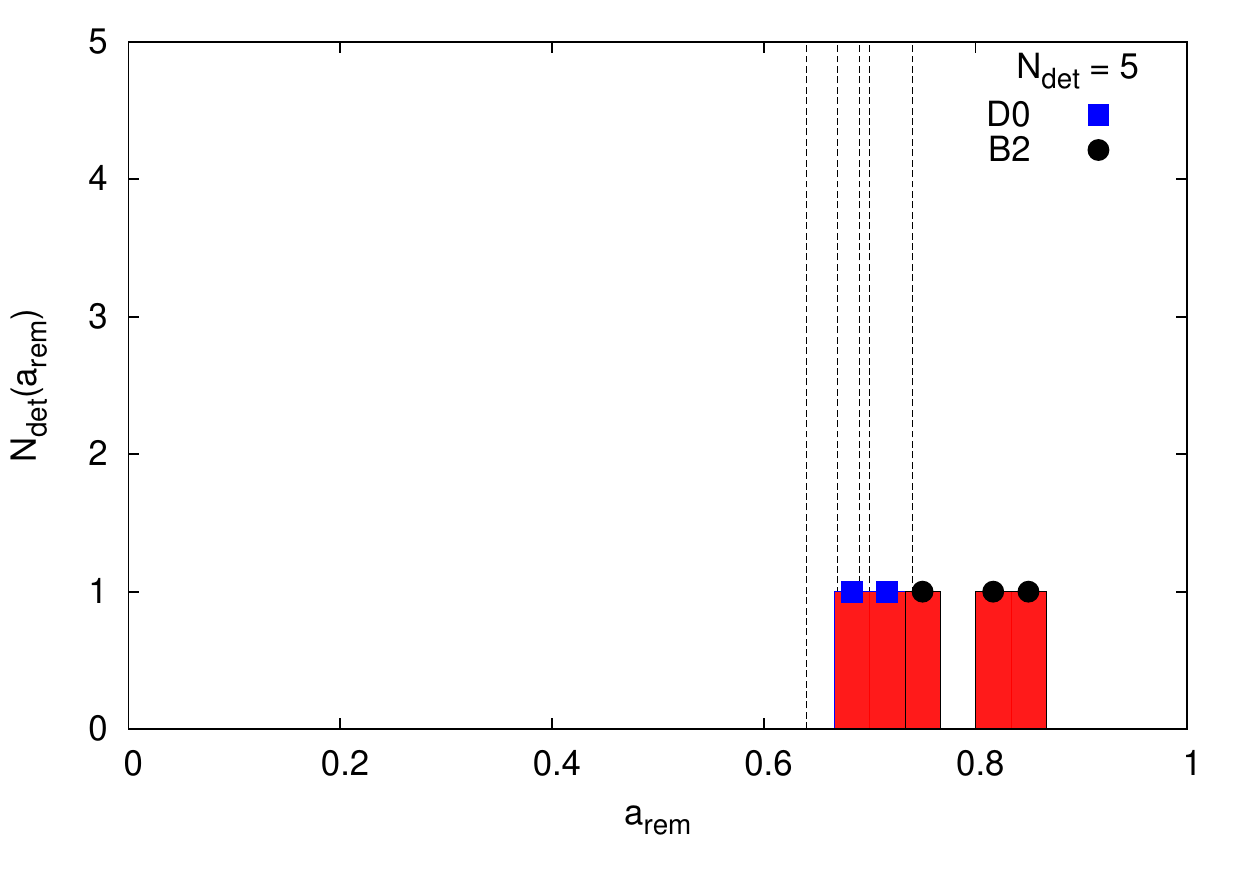}\\
\includegraphics[width=8cm]{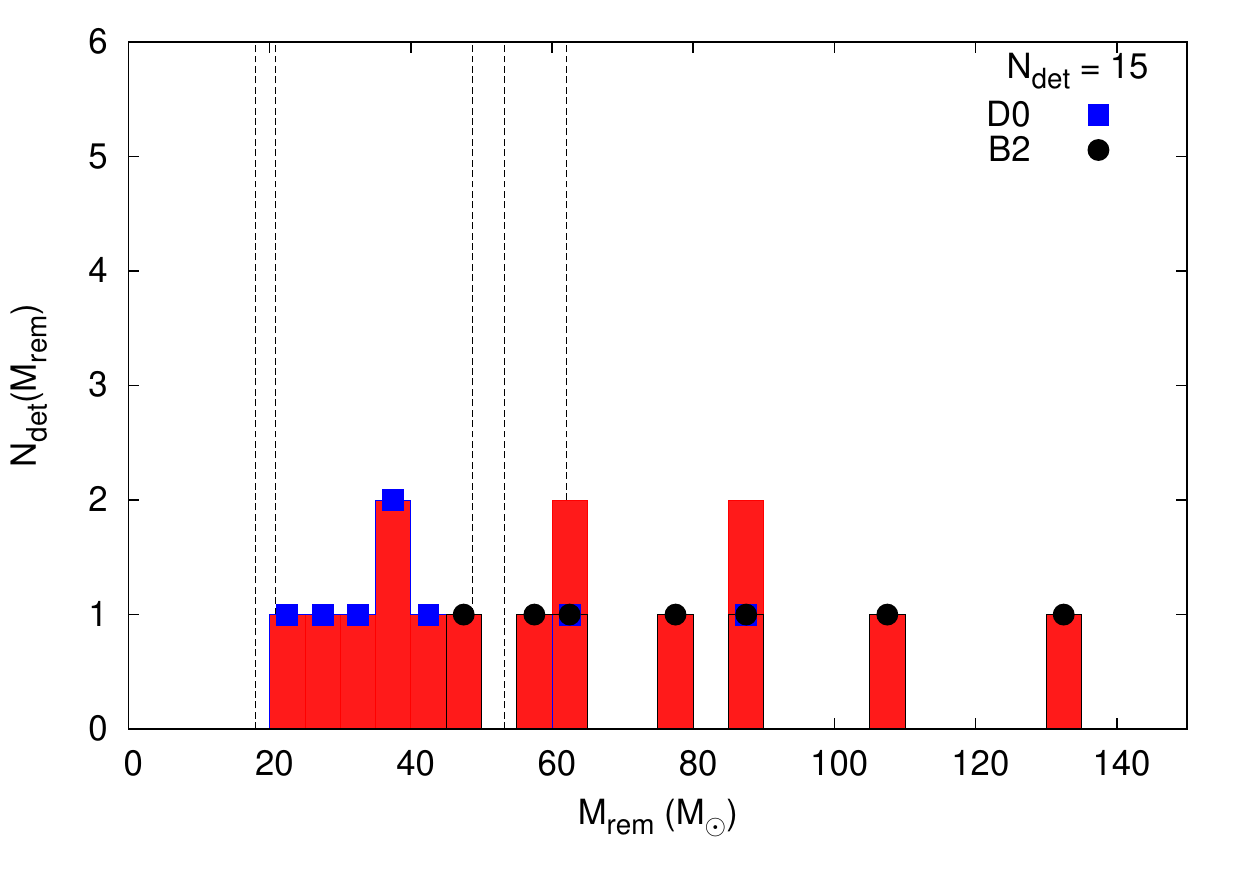}
\includegraphics[width=8cm]{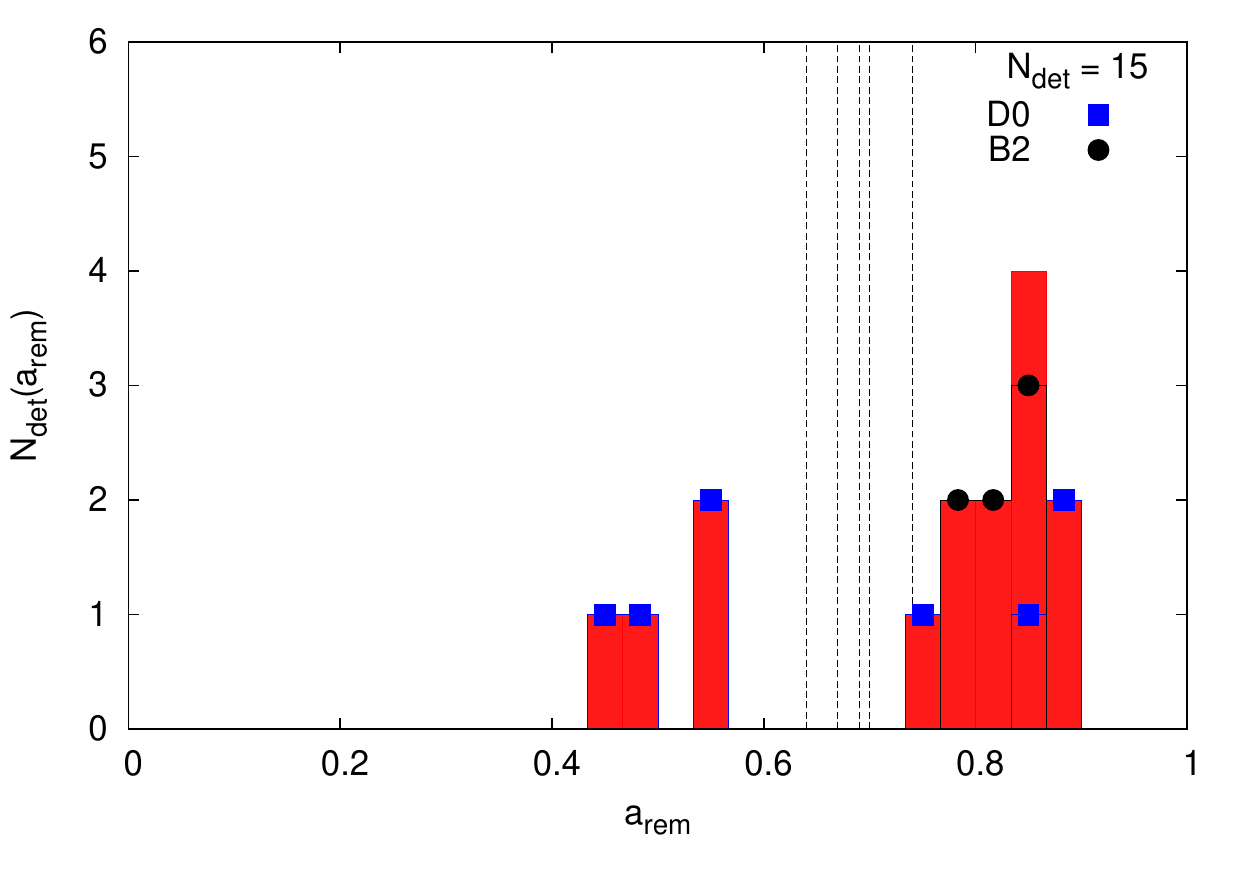}\\
\includegraphics[width=8cm]{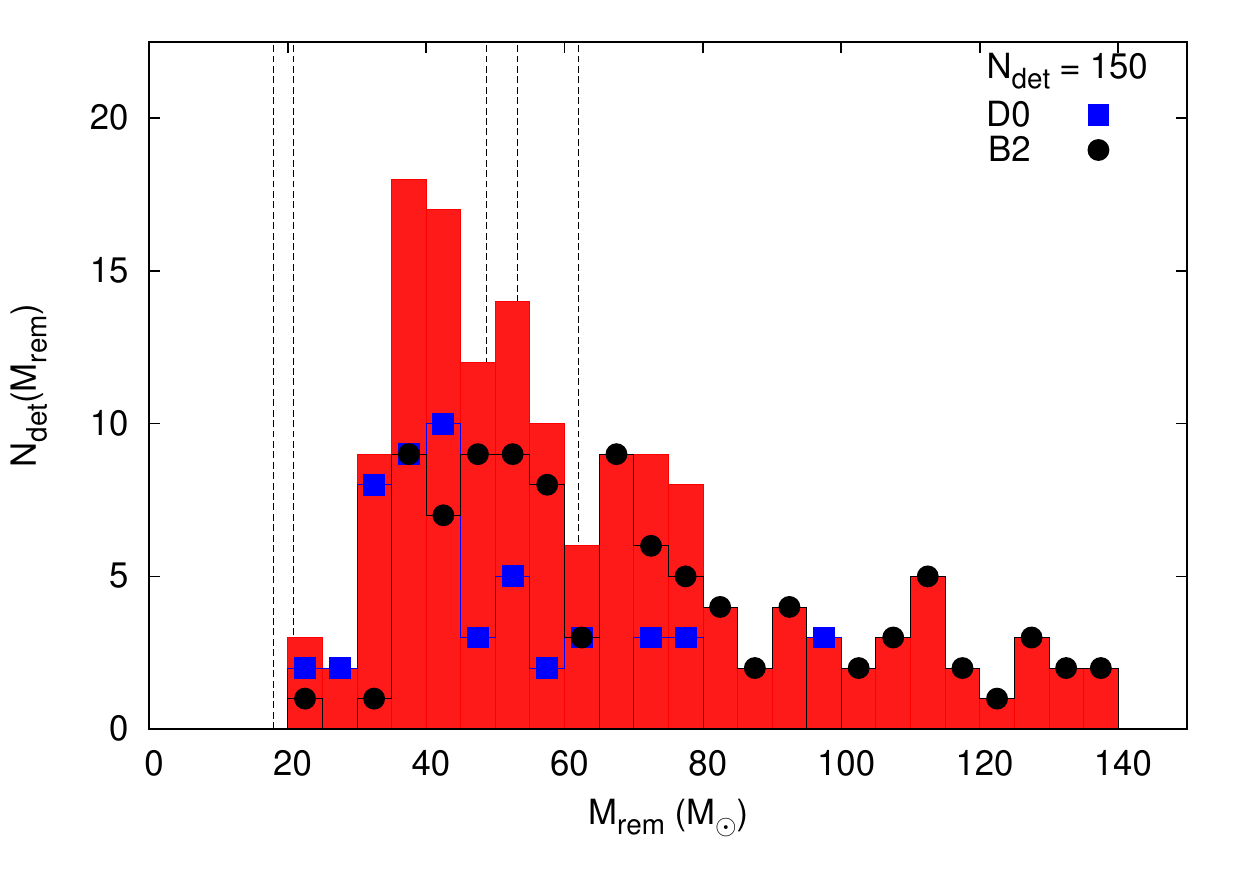}
\includegraphics[width=8cm]{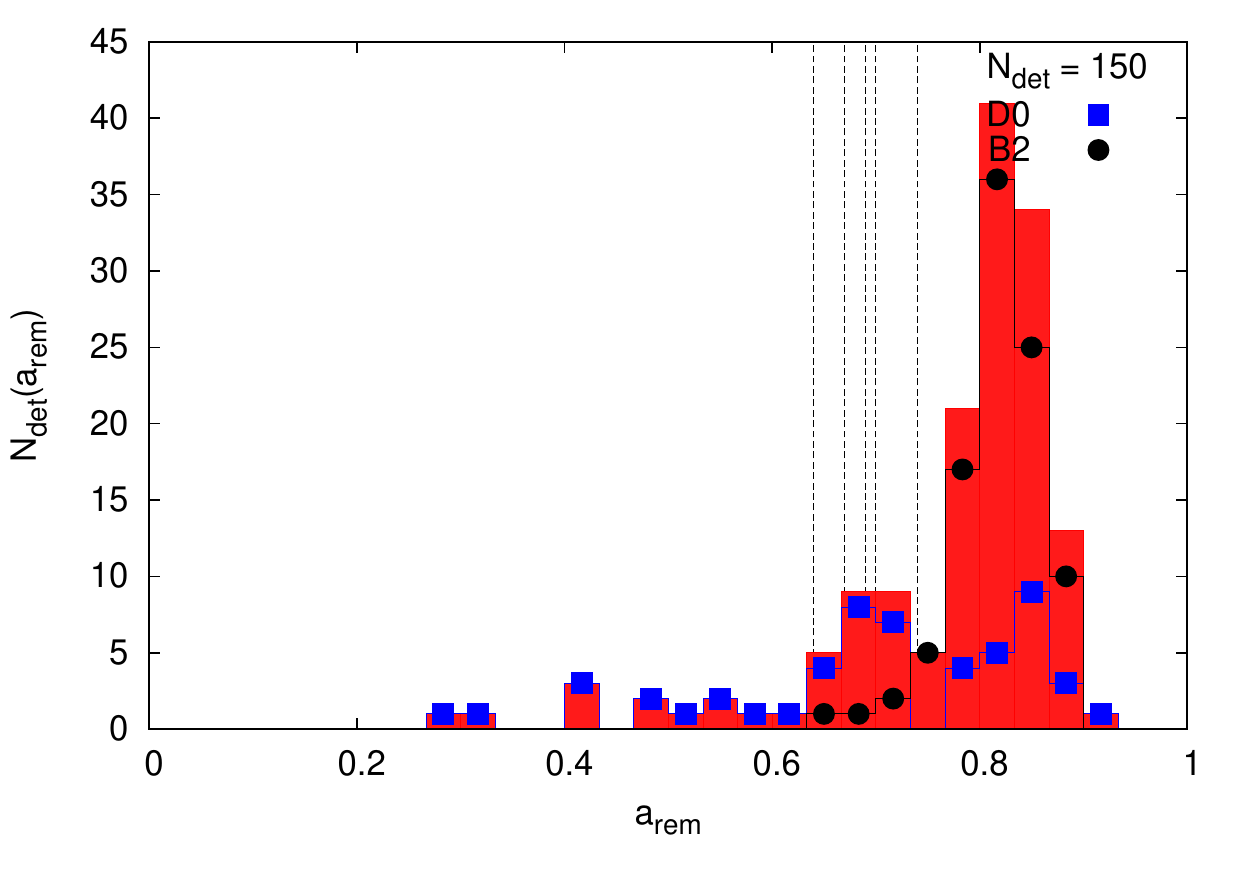}\\
\includegraphics[width=8cm]{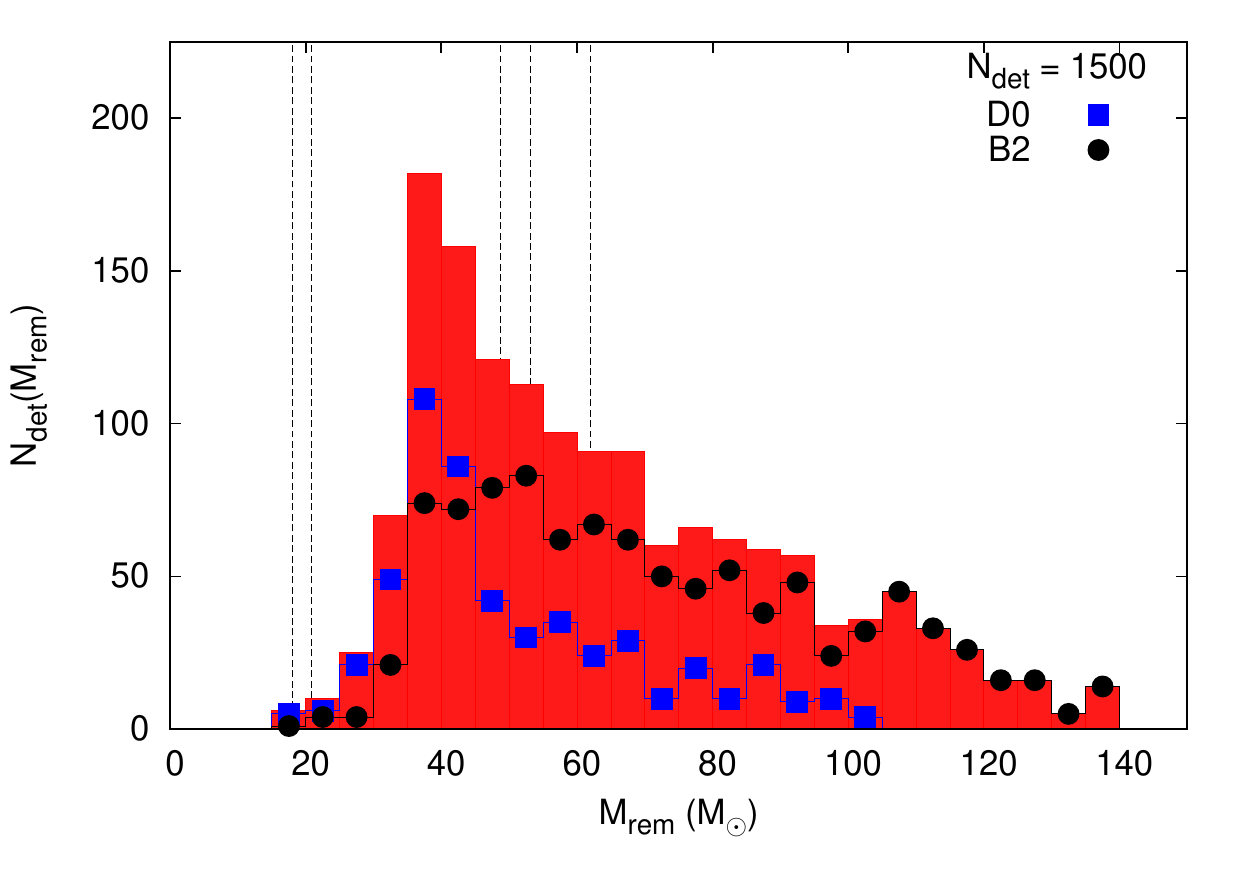}
\includegraphics[width=8cm]{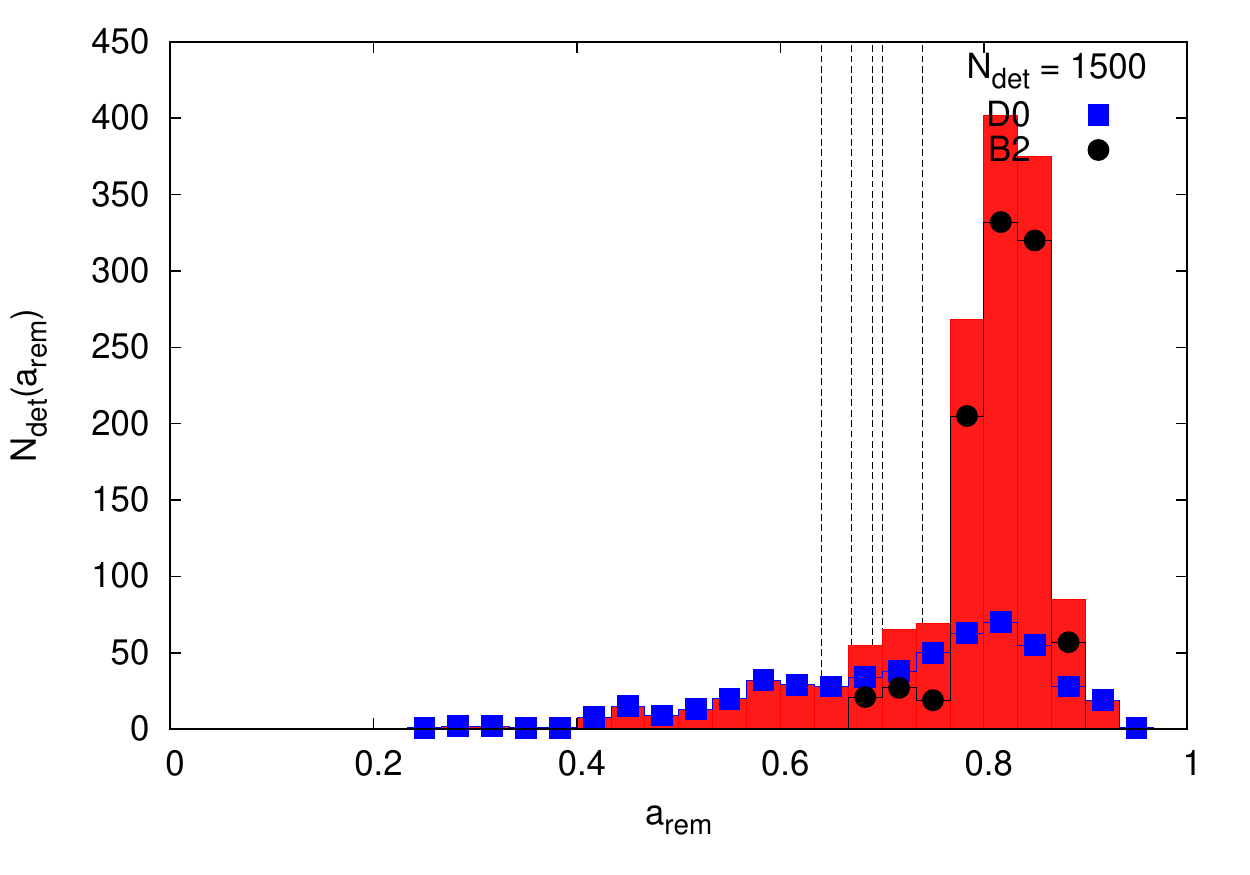}\\
\caption{Remnant mass (left column) and spin (right column) assuming, from top to bottom, a number of detection $N_{\rm det} = 5 ~- ~15 ~- ~50 ~- ~150$. The vertical dashed lines mark the currently observed population of BHs. In this case we are including the volume sensitivity dependence.}
\label{mmm}
\end{figure*}

According to the procedure described above, each BBH selected in this way represents a mock detection, thus providing a simple way to compare with LIGO detections.
Figure \ref{mmm} shows how the final BH mass and spin distributions vary at increasing the number of mock detections. 

For a low number of  detections, clearly the distribution is dominated by stochasticity, and we provide only one example that agrees nicely  with the presently observed population of BHs. 
 The picture remains unclear at detection numbers of order 10, as shown in the second row panels, although some separation starts to become evident, with dynamical mergers gathering at generally lower $M_\rem$ and $a_\rem$ values.

When the number of detections exceeds 100, disentangling the merged BH formation scenarios starts to become possible, especially in the $M_\rem$ high-end tail and in the $a_\rem$ low-end tail. 
However, clear signatures of different formation channels seem to emerge only when the detections number rises above $10^3$. Indeed, in this case it becomes evident that a simple dynamical scenario cannot account for BHs heavier than $100\Ms$, while misaligned isolated binaries cannot form BHs with spins below $0.6$.

In general, such an approach represents a powerful and flexible tool to predict what we expect from observations as soon as the detection number will increase. Comparing our models with observations can help in shedding light on different, still partly unknown, processes, like for instance the relation between the BHs natal spin and mass, the physics that regulate binary star evolution or the dynamical processes that drive the formation of BBHs in dense clusters.

However, we caution that the approach discussed here represents only the lowest order approximation, as a more careful investigation would deserve to include in this procedure other factors, like for instance different values of the stellar metallicity. 
In a companion paper, we will explore a wider region of the phase-space, with the aim on understanding how all the parameters discussed here can be used in concert to provide more detailed predictions.

\subsection{On the origin of observed BBH mergers}

In this section we use our models to infer likely formation scenarios for each of the currently observed population of merged BBHs.

\subsubsection*{GW150914}
The formation history of the heaviest LIGO source is the most difficult to disentangle. If BHs evolution is well described by the \SSEET stellar evolution scheme, it is quite probable that the GW150914 progenitor components were low-metal massive stars, with $Z\lesssim0.0002$, irrespective of the formation channel considered.
If the merger occurred in an isolated binary, its BHs were likely characterized by misaligned spins. Assuming \SSEET physics, this source represents the high-end tail of the mass distribution, as opposed to the \SSEVN mass distribution, for which it represents the low-end.
According to \SSEET results, a dynamical origin for GW150914 is possible if a small fraction of merged BHs were retained in the parent cluster and underwent a second merger event. Moreover, dropping any restriction on the BBH mass ratio leads to a larger number of BHs with masses in the range $40-60\Ms$, thus increasing the probability for GW150914-like sources to form.
On the other hand, using a different stellar evolution paradigm implies that the GW150914 progenitors must have a metallicity quite large $Z>0.002$ and the environment in which they evolved was not the site of a noticeable number of ``recycled'' mergers. 

We highlight here how important it is to improve our understanding of stellar evolution. Indeed, different stellar evolution recipes imply completely different origins for GW150914. If we believe  \SSEVN, such massive sources hardly form in low-metal environments and in isolated binaries. Low-density star clusters are the most probable site for their formation, with a BBH population characterised by a nearly flat mass ratio distribution and with an efficient retention fraction for merged BHs. If, instead, we believe to \SSEET recipes, then GW150914 likely formed in a metal-poor, massive and dense globular or nuclear cluster, where densitites are sufficiently high to retain merged BHs and recycle them in multiple merger events.

\subsubsection*{GW151226}

A clear outcome of our analysis is that GW151226 likely formed in an environment characterised by nearly solar metallicities. Both the isolated and dynamical channels provide a suitable explanation for its origin. If it formed in isolation, \SSEVN predictions allow us to conclude that GW151226 comes from a BBH with $Z=0.02$. Using \SSEET leads to similar conclusions, but a lower metallicity is needed to make  the remnant mass distribution and the GW151226 inferred mass coincide.
Concerning the spin orientation, it is unlikely that GW151226 originated from the merger of an aligned BBH. Indeed, in case of full alignment the spin distribution is extremely narrow and peaks at large spin values $\sim 0.85$, with little dependence on the metallicity. 
If it formed through dynamical interactions, the most probable host is a massive young cluster, characterised by solar metallicity and a capability of retaining a number of merged BHs. If this is the case, it is quite difficult to disentangle which stellar evolution recipes better explain its origin. In the case of a zero-recycling fraction, GW151226 barely falls within the most probable region of the $M_\rem-a_\rem$ plane. Indeed, assuming $f_\recy = 0$ we find that $M_\rem$ abruptly decreases at values above $25\Ms$.

\subsubsection*{GW170104}

This BH represents the current separation between ``light'' and ``heavy'' post merged BHs and its origin is quite difficult to assess. If its progenitor was an isolated binary, its evolution was characterised by an evident spin misalignment. The progenitor metallicity depends strongly on stellar evolution: \SSEET requires a quite low-metallicity $Z \simeq 2\times 10^{-4}$, while in the case of \SSEVN the GW170104 must have larger $Z$ values, ranging in between $2\times 10^{-3}-2\times 10^{-2}$. 

Moreover, there are a variety of configurations in the dynamical scenario that can explain the origin of GW170104. It falls well within the mass and spin distribution if it either formed in a low-density cluster, where the probability of recycling is low so BHs undergo only one merger, or in denser systems, where recycling is more likely. 
Again, stellar evolution is crucial in determining the progenitor metallicity: \SSEET predicts $Z \simeq 2\times 10^{-4}$, similar to the isolated channel, while \SSEVN leads to a metal-rich progenitor ($Z=0.02$) hosting recycled BHs. 

The best constraint on the origin of GW170104 is achieved by assuming an isolated binary progenitor characterized by randomly oriented spins and $Z = 0.0002$.

\subsubsection*{GW170608}

Similarly to GW151226, this source likely formed an environment richer in metals than other sources. If it formed in isolation, the progenitor metallicity was likely sub-solar, according to \SSEET. In the case of the dynamical scenario, instead, the most probable hosting cluster had solar metallicities and sufficiently high densities to host multiple series of BH mergers. Indeed, assuming $f_\recy=0$ leads GW170608 outside the $M_\rem-a_\rem$ most probable region.
Typical astrophysical systems having these properties are young massive clusters, characterized by high central densities, large metallicities and relatively small ages.
It is important to stress here that \SSEET does a better job of explaining the origin of GW170608, while with \SSEVN it falls systematically outside the $M_\rem$ full width at half maximum even assuming solar metallicities.
This would imply that GW170608 could have formed in a super-solar environment.

\subsubsection*{GW170814}

The formation of this source is well reproduced either by the isolated scenario, under the assumption of spin misalignment, or the dynamical scenario, assuming a small fraction of recycled BHs. \SSEET stellar evolution recipes point toward low-metallicities, $Z<0.002$, as opposed to \SSEVN, which suggest $0.002<Z<0.02$. 

A comparison between the GW170814 spins and our predictions seem to favour an isolated origin, provided that the progenitor spins were fully misaligned ($n_\theta = 0$).

\section{Conclusion}
In this paper we have explored the connections between the remnant mass and spin of merged BHs and the formation history of their progenitors. Using different stellar evolution codes, we investigated two evolutionary channels: the {\em isolated channel} in which BBHs form in isolation and merge after a common envelope phase, and the {\em dynamical channel} in which merging BBHs form through gravitational scatterings in a dense stellar cluster. We characterized the remnant mass and spin distributions for various evolutionary properties (progenitor metallicity, stellar evolution recipes) and environmental properties (spin misalignments, mass ratios, and recycling fraction). Our main results are summarized below.

Field binaries formed through the isolated channel are mostly affected by the level of misalignment between the component BH spin vectors and the binary orbital angular momentum. We describe the level of misalignment of each BH spin by $\theta$, the angle it makes with the orbital angular momentum vector. To mimic possible misalignments of the spins in the field BBH population, we distributed $\theta$ according to a probability function $\propto \left(\cos\theta + 1\right)^{n_\theta}$. The values of $n_\theta$ allow us to parametrize the misalignment level, $n_\theta = 0$ corresponds to a random spin orientation while $n_\theta \rightarrow \infty$ leads to a perfectly aligned population.

Our analysis suggests that a mild misalignment can reconcile some of the observed LIGO events with a field population. However, in the case of perfect alignment, the observed BHs fall in the least probably region of our modelled $M_{\rm rem}-a_{\rm rem}$ plane. When stellar evolution processes are taken into account self-consistently and a high level of misalignment is considered, the final spin calculated for low metallicity, $Z = 2 \times 10^{-4}$, is characterized by a broad distribution peaking at $a_\rem \simeq 0.7$, while the $M_\rem$ distribution shows a steep rise at masses $<40\Ms$ and a smooth decrease afterward. All the three most massive LIGO sources fall within this distribution.

The main parametrization for the dynamical binaries are the BBH mass ratio and the merging recycling fraction (i.e., the fraction of merged BHs that undergo a second merger event). We assumed that dynamical BBHs have a mass ratio that is either larger than a minimum value ($q_{\rm min}=0.5$) or that is unconstrained ($q_{\rm min} =0$). We also assumed that there was either no recycling or that $10\%$ of the population of merged BHs undergo another merger.

The choice of $q_{\rm min}=0.5$ leads to a tighter remnant mass distribution around a slightly smaller peak compared with the case of $q_{\rm min}=0$. Assuming a flat mass ratio distribution leads to $M_\rem$ which are generally smaller than $80\Ms$, while for $q_{\rm min}=0.5$ the $M_\rem$ distribution extends up to $\gtrsim 100\Ms$.
When a relatively small recycling fraction is introduced, the width of the $M_\rem$ distribution weakly increases. Both $f_\recy$ and $q_{\rm min}$ have only a marginal effect on the final remnant BH spin.

For low-metal systems, the $M_\rem$ distribution for dynamical BBHs is similar to the field population, with $M_\rem$ peaking at $\sim 40\Ms$, but the high mass tail is more prominent when a small recycling fraction is introduced.  The spin distribution is relatively broad, with a peak at $a_\rem = 0.8-0.85$.

Using a statistical approach based on the probability of detecting the merger of an isolated or a dynamically formed BBH, we show that a number of LIGO detections above $\sim 100$ will allow us to begin to see the fingerprints of the progenitor BBH formation history. Our results suggest that the high-end of the $M_\rem$ distribution will be dominated by merged BHs formed from isolated BBHs, while dynamically formed BBHs contribute mostly to the low-end tail of the $a_\rem$ distribution, especially in the region $a_\rem<0.6$.

The metallicity and stellar evolution recipes adopted are crucial in determining the final $M_\rem-a_\rem$ distribution. The updated \SSEET stellar evolution scheme suggests that the most massive BHs observed by LIGO were likely formed in a low-metal environment ($Z\simeq 2\times 10^{-4}$), while the lowest mass BHs originated from stars having solar-metallicities. Conversely, \SSEVN results suggest that even the most massive observed BHs formed in a sub-solar metallicity environment, likely having $2\times 10^{-3}\lesssim Z \lesssim 2\times 10^{-2}$.

Assuming \SSEET recipes, GW150914 might have originated in a metal poor dense star cluster. Its formation site possibly was a nuclear star cluster  or a massive globular cluster, probably characterized by a non-null recycling fraction. However, when using more recent stellar evolution recipes we found that the likely merging site changed significantly, being most probably a sub-solar metallicity star cluster with a density sufficiently low to avoid any noticeable recycling.

Both GW151226 and GW170608 likely formed in a solar-metallicity environment. If formed through dynamical interactions, a possible formation site is a massive open cluster. If formed in the field, the progenitor BBH was characterized by a large misalignment of the spin factors. Both the \SSEET and \SSEVN recipes lead to similar results. On another hand, using \BSEET seems very hard to produce solar-metallicity merging BBHs, thus suggesting that either they have formed through dynamical interactions, or another binary stellar evolution paradigm is needed.

The origin of GW170104 is more difficult to constrain since most of the configurations in this study can explain its properties. In other words, its final mass and spin are quite probable in both field and dynamical BBH, at both low and relatively large metallicities.

GW170814 most probably formed from a field binary with low metallicity, $Z = (0.2-2)\times 10^{-3}$, although a dynamical origin cannot be completely ruled out.

\section*{Acknowledgements}

The authors acknowledge the anonymous referee, whose comments have helped us in improving the content of this manuscript. The authors acknowledge also Albino Perego, David Keitel, Davide Gerosa and Christopher Berry for useful comments and discussion.
MAS acknowledges the Sonderforschungsbereich SFB 881 "The Milky Way System" (subproject Z2) of the German Research Foundation (DFG) for the financial support provided. MAS also acknowledges financial support from the Alexander von Humboldt Foundation and the Federal Ministry for Education and Research in the framework of the research project "The evolution of black holes from stellar to galactic scales". This work benefited from support by the International Space Science Institute, Bern, Switzerland, through its International Team programme ref. no. 393 "The Evolution of Rich Stellar Populations and BH Binaries" (2017-18).
MB acknowledges the support while serving at the National Science Foundation through award 1755085. Any opinion, findings, and conclusions or recommendations expressed in this material are those of the authors and do not necessarily reflect the views of the National Science Foundation. The authors thank the Nicolaus Copernicus Astronomical Center for the hospitality given during the development of part of this work.

\appendix

\section{Adapting SSE to SEVN}
\label{AppA}

Comparing with Figure 11 in \cite{spera15}, we found that the final BH masses can be easily obtained by the BH masses given by the SSE package through a linear approximation 
\begin{equation}
M_\bh = c_0M_{\rm SSE} + c1,
\end{equation}
provided that the $M_* - M_\bh$ plane is decomposed straightforwardly. For the sake of clarity, we provide in the following the parameters assuming a metallicity $Z = 0.002$:
\begin{equation}
(c_0,c_1) = 
\begin{cases}

(0.5,2.2)   & 18  <(M_*/\Ms)\leq 19.6,\\
(0.3,3)     & 19  <(M_*/\Ms)\leq 20,\\
(0.25,3.15) & 20  <(M_*/\Ms)\leq 21,\\
(0.5,1.1)   & 21  <(M_*/\Ms)\leq 23,\\
(0.6,-0.12) & 23  <(M_*/\Ms)\leq 25.6,\\
(10,-164)   & 25.6<(M_*/\Ms)\leq 26.3,\\
(35,-605.5) & 26.3<(M_*/\Ms)\leq 27.3,\\
(50,-870.5) & 27.3<(M_*/\Ms)\leq 28,\\
(-30,545.5) & 28  <(M_*/\Ms)\leq 28.5,\\
(-20,369)   & 28.5<(M_*/\Ms)\leq 29.2,\\
(-10,192.5) & 29.2<(M_*/\Ms)\leq 32.9,\\
(-1.1,47)   & 32.9<(M_*/\Ms)\leq 41.1,\\
(1.5,14)    & 41.1<(M_*/\Ms)\leq 53,\\
(2,5.5)     & 53  <(M_*/\Ms)\leq 60,\\
(-1.6,76.5) & 60  <(M_*/\Ms)\leq 64.8,\\
(4,-40.5)   & 64.8<(M_*/\Ms)\leq 70,\\
(-4,138.5)  & 70  <(M_*/\Ms)\leq 76,\\
(0.08,40)   & 76  <(M_*/\Ms)\leq 90,\\
(0.7,22.5)  & 90  <(M_*/\Ms)\leq 100,\\
(1.6,-5.5)  & 100 <(M_*/\Ms)\leq 150.
\end{cases}
\label{sevn}
\end{equation}

\section{The effect of BHs natal spin}
\label{appB}

As discussed in \cite{belczynski17}, Equation \ref{EQanat} displays the connection between the star properties before it turns into a BH and its natal spin. From Figure \ref{histo} it is clear that this relation implies that BHs with initial masses below $60\Ms$ form with spins relatively high $>0.6$. This translates in a remnant spin distribution peaking at values $0.7-0.85$, depending on the scenario investigated.

The currently observed population of BHs formed through BBH coalescence have spins in the range $0.6-0.8$, thus within the remnant spin distribution shown in our Figure \ref{histo}. At the same time, the potential future observation of low-spin BHs might represent an issue for our current understanding of how BHs form from dying stars. 
In order to understand how much the BH natal spins affect the results, we calculated the remnant mass and spin distribution for the D0 model, but assuming that BH natal spins do not depend on the BH natal mass. 

\begin{figure}
\centering
\includegraphics[width=8cm]{dynamical_D0}\\
\includegraphics[width=8cm]{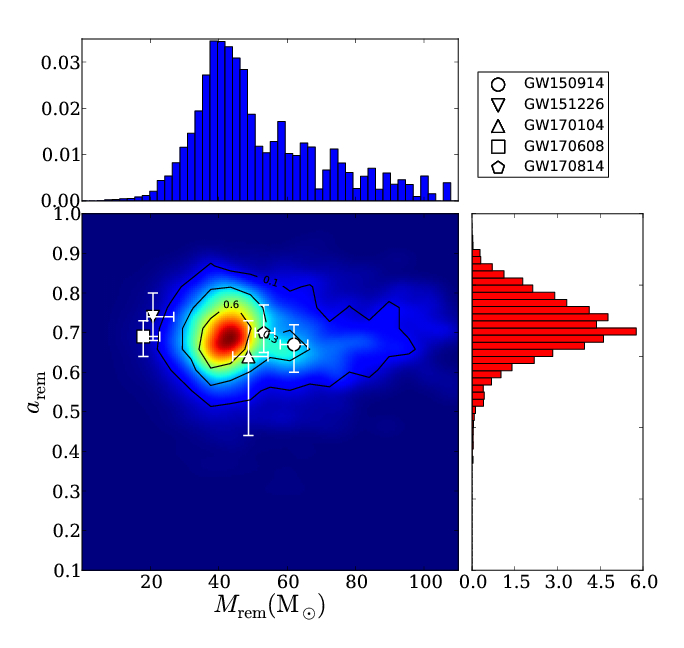}\\
\includegraphics[width=8cm]{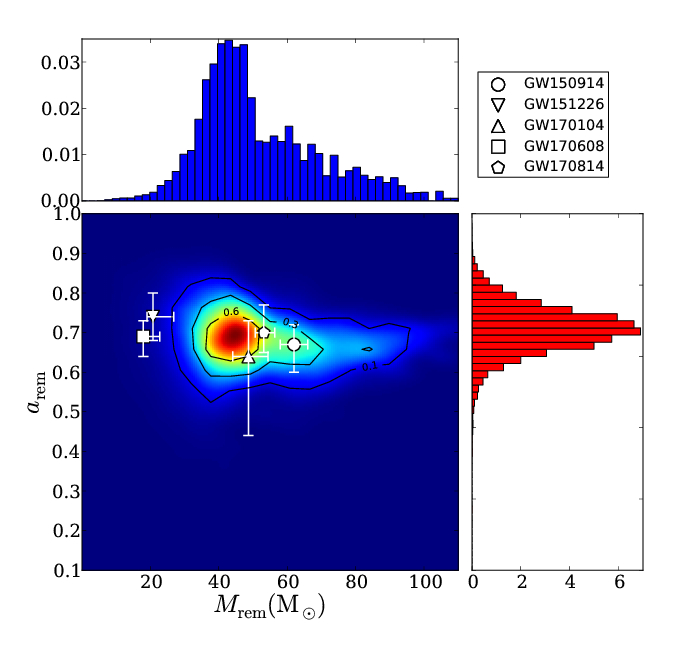}
\caption{
$M_\rem-a_\rem$ plane for model D0 (top panel), for the same model assuming a random distribution for BHs natal spins (central panel) or assuming a BH natal spin according to the \citet{amaro16} formula.
}
\label{appBf1}
\end{figure}

As shown in the central panel Figure \ref{appBf1}, assuming a random natal spin distribution significantly changes the remnant BH masses and spins, increasing the probability to form BHs with spins below $a_\rem \lesssim 0.6$, which are quite hard to produce assuming the natal spin-CO mass relation discussed above. Note the complete overlap between the model and the high-mass LIGO sources. These result suggest that only increasing the number of observed BHs would allow us to better discern whether our current knowledge of stellar BH formation is reliable or not.

Following \cite{amaro16}, we also investigated the case in which the BH natal spin is connected to the BH natal mass through the relation
\begin{equation}
a_\bh = (0.075 - 0.45)\left(\frac{M_\bh}{30\Ms}\right)^{-1}.
\end{equation}
For each BH in our sample, we extract a random value between the two limits given in the pre-factor in equation above to reproduce the same range of values as depicted in \cite{amaro16}.
The results obtained under this assumption, as shown in the bottom panel of Figure \ref{appBf1}, leads the $a_\rem$ distribution to be much narrower than in other cases, peaking at $a_\rem \sim 0.7$ and allowing the formation of BHs with low spin values, $a_\rem \sim 0.2$, though the probability associated to this objects is quite small.

\section{The effect of volume sensitivity}
\label{appC}
 
As discussed in Section \ref{results}, the volume sensitivity might affect the observed mass distribution of merged BHs. In the work presented here, we assumed the relation discussed by \cite{fishbach17}, which found $V\propto M_{\rm BBH}^{k}$, with $k \sim 2.2$ at fixed mass ratio and zero-spins. 

\begin{figure}
\centering
\includegraphics[width=8cm]{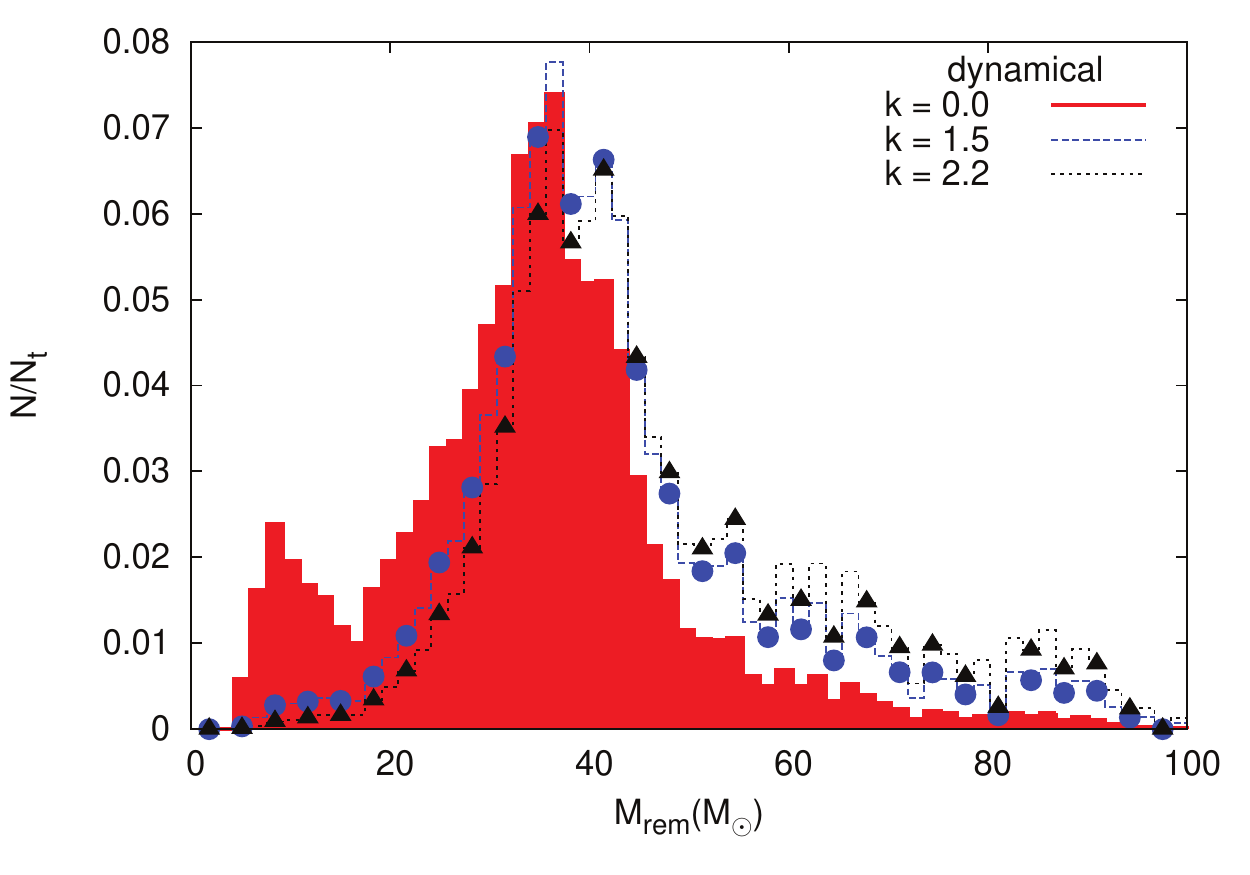}
\caption{
Mass distribution in model D0, weighted assuming a Volume - binary mass relation $V \propto M_{\rm BBH}^k$, for $k=0$ (red filled steps), $1.5$ (straight blue steps), $2.2$ (dotted black steps).
}
\label{Vcomp}
\end{figure}

However, mass ratio and the BBH components spin can affect significantly the sensitive volume, potentially affecting also the dependence on the total BBH mass. In order to understand how much our results would change at varying the slope of the $V-M_{\rm BBH}$ relation, we show in Figure \ref{Vcomp} a comparison in the D0 model, assuming either $k = 2.2$, $k=1.5$ or $k=0$. 
The major effect that the $V-M_{\rm BBH}$ relation has on the BBH mass distribution is to cause the abrupt decrease of the low-end tail, thus causing a magnification of the probability to observe BHs with masses above $\sim 50 \Ms$. Therefore, a slower slope can shift the distribution peak to lower values, possibly leading the results obtained assuming more recent stellar evolution recipes (obtained with the \SSEVN code) to a range compatible with the LIGO sources population.
This highlight the importance of providing a comprehensive study based on how the sensitive volume varies at varying BBH mass and components mass and spin, all quantities that can sensitively affect the volume-mass relation.

Another quantity that in principle might affect the $M_\rem$ distribution is the energy radiated during the merger, $E_{\rm rad} = 1-M_\rem/(M_1+M_2)$, which depends on the BBH components spins and masses through the remnant mass. 
This parameter allows to take into account the fact that LIGO's sensitivity is larger for aligned spins BBHs compared to non-spinning systems, and that unequal-mass BBHs are harder to be observed compared to equal-mass BBHs since they radiate more energy. 

For the sake of completeness, we also try to scale the $M_\rem$ distribution assuming that the LIGO's sensitive volume is linked to the radiated energy via a power-law. However, we find that $E_{\rm rad}$ affects all the mass bins in the same way, on average, since it depends only on the BBH mass ratio and spins, thus leading to a little-to-no effect on the final $M_\rem$ distribution.

\clearpage
\footnotesize{
\bibliographystyle{mnras}
\bibliography{biblio}
}

\end{document}